\documentclass[aip,jcp,reprint]{revtex4-2}
\usepackage[utf8]{inputenc} 
\usepackage[T1]{fontenc}       
\usepackage{amssymb}
\usepackage{amsmath}
\usepackage{comment} 

\usepackage{float} 
\usepackage{graphicx} 
\usepackage{color}
\usepackage[singlelinecheck=false,justification=raggedright]{caption}
\usepackage[parfill]{parskip} 

\newcommand{\mean}[1]{\langle #1 \rangle}

\makeatletter
\newcommand*{\rom}[1]{\expandafter\@slowromancap\romannumeral #1@}
\makeatother

\begin{document}
\title{Stochastic Real-Time Second-Order Green's Function Theory for Neutral Excitations in Molecules and Nanostructures} 

\author{Leopoldo Mej\'ia}
\email{leopoldo.mejia@berkeley.edu}
\affiliation{Department of Chemistry, University of California, Berkeley, California 94720, USA}
\affiliation{Materials Sciences Division, Lawrence Berkeley National Laboratory, Berkeley, California 94720, USA}

\author{Jia Yin}
\email{jiayin@lbl.gov}
\affiliation{Applied Mathematics and Computational Research Division, Lawrence Berkeley National Laboratory, Berkeley, California 94720, USA}

\author{David R. Reichman}
\email{drr2103@columbia.edu}
\affiliation{Department of Chemistry, Columbia University, New York, New York 10027, USA}

\author{Roi Baer}
\email{roi.baer@huji.ac.il}
\affiliation{Fritz Haber Center for Molecular Dynamics, Institute of Chemistry, The Hebrew University of Jerusalem, Jerusalem 91904, Israel}

\author{Chao Yang}
\email{cyang@lbl.gov}
\affiliation{Applied Mathematics and Computational Research Division, Lawrence Berkeley National Laboratory, Berkeley, California 94720, USA}

\author{Eran Rabani}
\email{eran.rabani@berkeley.edu}
\affiliation{Department of Chemistry, University of California, Berkeley, California 94720, USA}
\affiliation{Materials Sciences Division, Lawrence Berkeley National Laboratory, Berkeley, California 94720, USA}
\affiliation{The Raymond and Beverly Sackler Center of Computational Molecular and Materials Science, Tel Aviv University, Tel Aviv 69978, Israel}

\begin{abstract}
We present a real-time second-order Green's function (GF) method for computing excited states in molecules and nanostructures, with a computational scaling of $O(N_{\rm e}^3$), where $N_{\rm e}$ is the number of electrons.  The cubic scaling is achieved by adopting the stochastic resolution of the identity to decouple the 4-index electron repulsion integrals (ERI). To improve the time-propagation and the spectral resolution, we adopt the dynamic mode decomposition (DMD) technique and assess the accuracy and efficiency of the combined approach for a chain of hydrogen dimer molecules of different lengths. We find that the stochastic implementation accurately reproduces the deterministic results for the electronic dynamics and excitation energies. Furthermore, we provide a detailed analysis of the statistical errors, bias, and long-time extrapolation. Overall, the approach offers an efficient route to investigate excited states in extended systems with open or closed boundary conditions.
\end{abstract}

\maketitle

\section{\label{sec:intro}Introduction}
The computation of excited state properties is a very active field in the molecular and materials sciences.\cite{adamo2013calculations, gonzalez2012progress, lischka2018multireference, dreuw2015algebraic, hernandez2018first, doutime2022, jasrasaria2020sub, del2020accelerating, liang2022revisiting, hait2021orbital, higgott2019variational, faber2014excited} The importance of such calculations is accentuated by the wide range of technological applications that are derived from a deeper understanding of excited state properties, as well as the fundamental physics and chemistry that can be learned from the development of methods to compute them. In molecular systems, time-dependent density functional theory~\cite{marques2004time, burke2005time, casida2012progress, marques2006time2} (TDDFT) or wave function-based methods, such as time-dependent Hartree-Fock~\cite{mclachlan1964time, jorgensen1975molecular, li2005time} (TDHF) and coupled cluster within the equation of motion formalism (EOM-CC),\cite{hirata2004higher, bartlett2012coupled} are commonly used to compute excited state energies. However, it is challenging to find a balance between accuracy and efficiency. While methods such as TDDFT and TDHF can handle the computation of the excited state properties of systems containing hundreds of electrons, their accuracy highly depends on the system or system-functional combination, in the case of TDDFT. By contrast, while wave-function-based methods that include electron correlation beyond the level of Hartree-Fock (e.g. EOM-CC) are usually more accurate, their inherent steep computational cost restricts computations to systems with a few atoms only.\cite{krylov2008equation, hirata2004higher, bartlett2012coupled}

Alternative methods traditionally used in condensed matter theory, such as many-body perturbation theory (MBPT) within the Green's function (GF) formalism,\cite{stefanucci2013non, economou2006green, kadanoff2018quantum} have also proven to be useful to describe excited states. Two of the most popular approximations are the GW method~\cite{hybertsen1986electron, aryasetiawan1998gw, van2006quasiparticle, kotani2007quasiparticle, golze2019gw, shishkin2007self}, a first-order approximation to the self-energy in the {\em screened} Coulomb interaction ($W$) and the GF2 method,\cite{phillips2014communication, pavovsevic2017communication} in which the self-energy is approximated to second-order in the {\em bare} Coulomb interaction, allowing for the inclusion of dynamical exchange correlations. The GW and the GF2 closures have been successfully used to compute charged excitations (quasiparticle energies) in molecules and bulk systems~\cite{hybertsen1986electron, dou2019stochastic, van2015gw, caruso2016benchmark} and have been extended to describe neutral excitations using time-dependent approaches.\cite{rohlfing2000electron, doutime2022} 
Attaccalite~{\em et al.}~\cite{attaccalite2011real} showed that the time-dependent GW approach is equivalent to the well-known Bethe-Salpeter equation (BSE) in the adiabatic, linear response limit.  Similarly, Dou {\em et al.}\cite{doutime2022} derived a Bethe-Salpeter-like equation with a second-order kernel (GF2-BSE) and tested the approach for a set of molecules, with encouraging results for low-lying excited states, particularly for charge transfer excitations.\cite{doutime2022} However, the $O(N_{\rm e}^6)$ scaling with system size (or $O(N_{\rm e}^5)$ in real-time), where $N$ is the number of electrons, of both GW and GF2 approaches limit their applications to relatively small systems or basis set sizes.

Here, we develop a stochastic real-time approach to obtain neutral excitations within the second-order Born approximation (GF2), reducing the computational scaling from $O_{\rm e}^6)$ to $O(N_{\rm e}^3)$. This is achieved using the range-separated~\cite{dou2020range} stochastic resolution of the identity~\cite{takeshita2017stochastic} to decouple the 4-index electron repulsion integrals appearing in the Kadanoff-Baym (KB) equations.\cite{kadanoff2018quantum} Furthermore, we adopt the dynamic mode decomposition (DMD) technique~\cite{kutz2016dynamic,yin2023analyzing, yin2022using, reeves2022dynamic} to solve the nonlinear Kadanoff-Baym equations within the adiabatic approximation. The DMD method is a data-driven model order reduction procedure used to predict the long-time nonlinear dynamics of high-dimensional systems and has been used previously with the time-dependent GW approach.\cite{reeves2022dynamic} We assess the accuracy of the stochastic, real-time GF2 approach with respect to the number of stochastic orbitals, the propagation time, and the system size for hydrogen dimer chains of varying lengths.

The manuscript is organized as follows. In Sec.~\ref{sec:theory} and Sec.~\ref{sec:stc-TD}, we summarize the GF2-BSE method and introduce the stochastic approaches to its real-time implementation, respectively. In Sec.~\ref{sec:results}, we compare the real-time stochastic and deterministic algorithms, analyze the statistical error in the computations, and evaluate the quality of the DMD extrapolation. Finally, in Sec.~\ref{sec:conclusions} we discuss the significance and perspectives of this work.

\section{Time-dependent GF2}
\label{sec:theory}
In this section, we provide a summary of the time-dependent GF2 approach for computing neutral excitations.\cite{doutime2022} We begin by defining the electronic Hamiltonian in second quantization. Next, we summarize the Kadanoff-Baym equations (KBEs) for the two-time GF on the Keldysh contour and introduce the second-order Born approximation. Finally, we describe the adiabatic limit to the KBEs. 

\subsection{Hamiltonian}
We consider the electronic Hamiltonian of a finite system interacting with an explicit electric field. In second quantization the Hamiltonian is given by
\begin{equation}
    \hat H = \sum_{ij} h_{ij}   \hat a^\dagger_i \hat a_j + \frac12 \sum_{ijkl}v_{ijkl}  \hat a^\dagger_i \hat a^\dagger_k  \hat a_l \hat a_j + \sum_{ij} \Delta_{ij}(t) \hat a^\dagger_i \hat a_j~,
    \label{eq:H}
\end{equation}
where $i$, $j$, $k$, and $l$ denote indexes of a general basis, $\hat{a}^\dagger_i$ ($\hat{a}_i$) is the creation (annihilation) operator for an electron in orbital $\chi_i$, and $h_{ij}$ and $v_{ijkl}$ are the matrix elements of the one-body and two-body interactions, respectively. The two-body terms are given by the 4-index electron repulsion integral (ERI):
\begin{equation} 
	\label{eq:2e4c}
	v_{ijkl} = (i  j | kl ) = \iint  \frac{\chi_i ({\bf r}_1)\chi_j ({\bf r}_1)\chi_k ({\bf r}_2)\chi_l ({\bf r}_2)}{\left|{\bf r}_1-{\bf r}_2\right|} d{\bf r}_1 d{\bf r}_2,
\end{equation}
where we have assumed that the basis set is real. We use atomic units throughout the manuscript, where the electron charge $e=1$, the electron mass $m_e=1$, $\hbar=1$, the Bohr radius $a_0=1$, and $4\pi\epsilon_0=1$. 

The last term in Eq.~\eqref{eq:H} is a time-dependent perturbation. Here, to describe neutral excitation, we couple the system to an external electric field, $E(t)$, within the dipole approximation, where $\Delta_{ij}(t)=E(t) \cdot {\bf \mu}_{ij}$ and 
\begin{equation}
 {\bf \mu}_{ij}=\int \chi_i({\bf r}) {\bf r} \chi_j({\bf r}) d{\bf r}.
 \label{eq:mu_ij}
\end{equation}
We choose to explicitly include this term in the Hamiltonian (rather than introducing a linear-response perturbation in the initial wave function) because we make no assumption that the field is weak.

\subsection{Kadanoff-Baym Equations}
Following Ref.~\onlinecite{doutime2022}, the equations of motion for the single-particle lesser Green's function, ${\bf G}^<(t_1,t_2)$, are given by the KB equations:
\begin{equation}
    i\partial_{t_1}{\bf G}^<(t_1,t_2) = {\bf F}[\rho(t_1)]{\bf G}^<(t_1,t_2) + {\bf I}^<_\alpha(t_1,t_2)
    \label{eq:kb1}
\end{equation}
and
\begin{equation}
    -i\partial_{t_2}{\bf G}^<(t_1,t_2) = {\bf G}^<(t_1,t_2){\bf F}[\rho(t_2)] + {\bf I}^<_\beta(t_1,t_2)~,
    \label{eq:kb2}
\end{equation}
where $t_1$ and $t_2$ are projections onto the real-time branch, $\rho(t)=-i {\bf G}^<(t, t)$ is the time-dependent density matrix, and ${\bf F}[\rho(t)]$ is the Fock operator, with matrix elements given by
\begin{equation}
    F_{ij}[\rho(t)] = h_{ij} + v^H_{ij}[\rho(t)] + v_{ij}^{x} [\rho(t)] + \Delta_{ij}(t).
\end{equation}
In the above, the Hartree and exchange potentials are given by $v^H_{ij}[\rho]=\sum_{kl}v_{ijkl}\rho_{kl}$  and $v_{ij}^{x} [\rho]=\sum_{kl}v_{ikjl}\rho_{kl}$, respectively. 

The last terms in Eqs.~\eqref{eq:kb1} and \eqref{eq:kb2} are the collision integrals, given by\cite{doutime2022}
\begin{equation}
  \label{eq:Ia}
    \begin{split}
    {\bf I}^<_\alpha(t_1,t_2) =& \int_0^{t_1} {\bf \Sigma}^R(t_1,t_3){\bf G}^<(t_3,t_1)dt_3\\
    &+ \int_0^{t_2} {\bf \Sigma}^<(t_1,t_3){\bf G}^A(t_3,t_1)dt_3
    \end{split}
\end{equation}
and
\begin{equation}
\label{eq:Ib}
    \begin{split}
    {\bf I}^<_\beta(t_1,t_2) =& \int_0^{t_1} {\bf G}^R(t_1,t_3){\bf \Sigma}^<(t_3,t_2)dt_3\\
    &+ \int_0^{t_2} {\bf G}^<(t_1,t_3){\bf \Sigma}^A(t_3,t_2)dt_3~,
    \end{split}
\end{equation}
respectively. In the above equations,  ${\bf \Sigma}$ is the self-energy (which encodes all many-body interactions) and the superscript "$R / A$" denotes retarded/advanced components.

\subsection{Second-Order Born Approximation to the Self-Energy}
\begin{figure}[t]
 \centering
\includegraphics[scale=1.0]{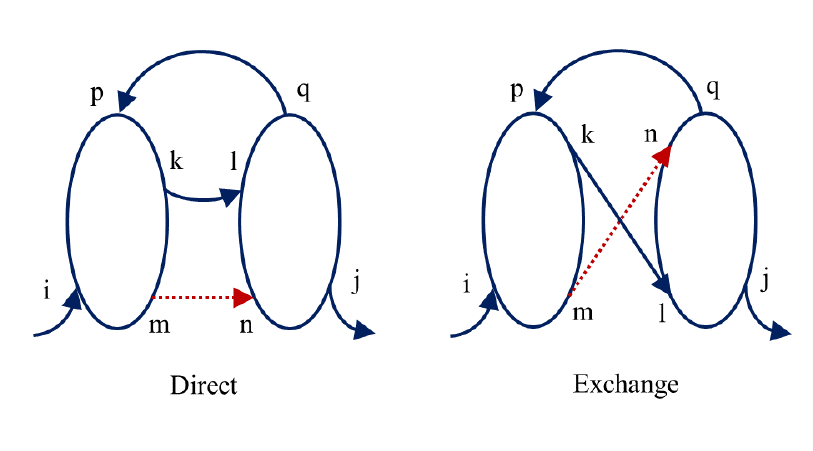}
\caption{Direct and exchange correlations contained in the Second-order Born self-energy, GF2. The ovals represent electron repulsion integrals and the arrows are propagators (Green's functions). In Eq.~\eqref{eq:self-GF2}, the blue components of the diagrams (solid lines) are wrapped into the screened Coulomb interaction, while the red propagator ($m\to n$, dotted arrows) is explicitly kept as $G_{mn}$.}
\label{fig:GF2}
\end{figure}

To obtain an approximate expression for the self-energies, we use the second-order Born approximation, where the self-energy is expanded to second-order in the Coulomb interaction. The resulting retarded component can be written in terms of the retarded and greater screened Coulomb integrals ($\delta W^{R/>}$)\cite{doutime2022}
\begin{equation}
    \begin{split}
    \Sigma^R_{ij}(t_1, t_2 ) =& \sum_{mn} i G^<_{mn}(t_1, t_2 ) \delta W^R_{imjn}(t_1, t_2 ) \\ 
    &+i G^R_{mn}(t_1, t_2 )\delta W^>_{imjn}(t_1, t_2 )~,
    \end{split}
    \label{eq:self-GF2}
\end{equation}
where
\begin{equation}
    \begin{split}
    \delta W&^R_{imjn}(t_1, t_2 ) = -i  \sum_{klqp}   ( G^<_{kl}(t_1, t_2 ) G^A_{qp}(t_2, t_1 )\\  
    &+ G^R_{kl}(t_1, t_2 ) G^<_{qp}(t_2, t_1 )  )v_{impk} (2 v_{jnql} - v_{jlqn} )
    \end{split}
\end{equation}
and
\begin{equation}
\begin{split}
\delta W&^>_{imjn}(t_1, t_2 ) =\\ &-i  \sum_{klqp}  G^>_{kl}(t_1, t_2 ) G^<_{qp}(t_2, t_1 ) v_{impk} (2 v_{jnql} - v_{jlqn} ).
\end{split}
\end{equation}
A particular feature of the self-energy in Eq.~\eqref{eq:self-GF2} is the inclusion of dynamical exchange as diagrammatically illustrated in Fig.~\ref{fig:GF2}.

\subsection{The Adiabatic Approximation}
The equations of motion \eqref{eq:kb1} and \eqref{eq:kb2} for the GFs together with the expression for the self-energy (Eq.~\eqref{eq:self-GF2}) form a close set of equations, but depend on two times, $t_1$ and $t_2$. To further simplify the time evolution of the GF, we assume that the retarded self-energy responds instantaneously to the application of external driving forces (e.g. the adiabatic limit)\cite{doutime2022}
\begin{equation}
\label{eq:selfad}
    {\bf \Sigma}^R(t_1,t_2) \approx \tilde{\bf \Sigma}^{\rm ad}[(t_1+t_2)/2]\delta(t_1-t_2)~,
\end{equation}
while the lesser self-energy is assumed to be negligible\cite{doutime2022}
\begin{equation}
\label{eq:selflessad}
    {\bf \Sigma}^<(t_1,t_2) \approx 0~.
\end{equation}
In the above, $\tilde{\bf \Sigma}^{\rm ad}[t]$ is the so-called adiabatic GF2 self-energy with matrix elements\cite{doutime2022}
\begin{equation}
\label{eq:selfad_final}
    \tilde{\Sigma}^{\rm ad}_{ij}(t) = -\sum_{mn}\delta\tilde{W}^{R}_{imjn}\rho_{mn}(t)
    +\frac{1}{2}\Re\sum_{mn}\delta\tilde{W}^{R}_{imjn}\delta_{mn}~,
\end{equation}
where
\begin{widetext}
\begin{equation}
    \delta\tilde{W}^{R}_{imjn} =\lim_{\omega \to 0}\left\{-\frac{1}{2}\sum_{kq} \frac{f(\varepsilon_k)-f(\varepsilon_q)} {\varepsilon_k-\omega-\varepsilon_q-i\eta} v_{imqk}(2v_{jnqk}-v_{jkqn})\right\}
\label{eq:deltaW}    
\end{equation}
\end{widetext}
is the Fourier transform of the screened Coulomb interaction, $f(\varepsilon)$ is the Fermi-Dirac distribution, $\eta$ is a small positive regularization parameter, and $\varepsilon_k$ are the quasiparticle energies obtained using a stochastic GF2 for charge excitations (see Ref.~\onlinecite{dou2019stochastic} for more information on how to calculate the quasiparticle energies using GF2). Using Eqs.~\eqref{eq:selfad} and \eqref{eq:selflessad} for the self-energy, the time evolution of the GFs given by Eqs.~\eqref{eq:kb1} and \eqref{eq:kb2} can be reduced to a simpler form for the equal time ($t_1 = t_2\equiv t$) GF (we assume an orthogonal basis from now on)\cite{doutime2022}
\begin{equation}
\label{eq:rhoEOM}
    \begin{split}
    i\frac{d}{dt}\rho(t) = [{\bf F}[\rho(t)],\rho(t)] + \tilde{\bf \Sigma}^{ad}(t)\rho(t) - \rho(t)\tilde{\bf \Sigma}^{ad\dag}(t),
     \end{split}
\end{equation}
where, as before, $\rho(t)=-i{\bf G}^<(t,t)$ and $[A,B] = AB-BA$. Excitation energies obtained using Eq.~\eqref{eq:rhoEOM} will be referred to as TD-GF2 (or TD-G0F2 when the quasi-particle energies are corrected using a single-shot, non-self-consistent GF2~\cite{dou2019stochastic}). 

In TD-GF2, the computational limiting step is the calculation of the self-energy at time $t$, $\tilde{\bf \Sigma}^{ad}(t)$. The formal computational cost scales as $O(N_e^5)$ with system size, limiting the application of TD-GF2 to small system sizes.  To reduce the number of self-energy evaluations, we adopt the dynamic mode decomposition (DMD) method to describe the long-time limit of the density matrix, $\rho(t)$, as described in the next subsection. In addition, we develop a stochastic approach that reduces the scaling of computing the self-energy to $O(N_e^3)$ at the account of introducing a controlled statistical error. This is described in the next section.

\subsection{Dynamic Mode Decomposition}
The dynamic mode decomposition method allows the extrapolation of the density matrix dynamics to long times without the need to further solve the equation of motion. As developed in Ref. \onlinecite{yin2023analyzing}, the DMD method is a data-driven model order reduction procedure used to predict the long-time nonlinear dynamics of high-dimensional systems. The method is based on Koopman's theory~\cite{koopman1931ham, koopman1932dyn} for reduced order modeling. The general strategy is to find a few ($r$) modes $\phi_{\ell}^{ij}$ with associated frequencies $\omega^{\ell}_{ij}$ to approximate the density matrix dynamics as
\begin{equation}
\label{eq:DMD}
    \rho_{ij}(t) = \sum_{l=1}^r\lambda_{ij} ^{\ell}\phi^{\ell}_{ij} e^{i\omega^{\ell}_{ij} t}
\end{equation}
with coefficients $\lambda^{\ell}_{ij}$. This model is constructed from the short-time nonlinear dynamics of the density matrix and can be seen as a finite-dimensional linear approximation to the dynamics.

\section{Stochastic Real-time GF2 Approach}
\label{sec:stc-TD}
In this section, we adopt the stochastic resolution of the identity~\cite{takeshita2017stochastic,dou2020range} to calculate the adiabatic self-energy appearing in Eq.~\eqref{eq:selfad_final} and combine it with the equation of motion for the density matrix (cf., Eq.~\eqref{eq:rhoEOM}).

\subsection{Stochastic Vectors and the Resolution of the Identity}
We define a \emph{stochastic orbital} $\theta$ as a vector in the Hilbert space of the system with random elements $\pm 1$. The average of the outer product of the stochastic vectors
\begin{equation}
\label{eq:stcI}
    \mean{ \theta\otimes\theta^T}_{N_s\to\infty} =
     \begin{pmatrix}
    1 & 0 & \cdots & 0\\
    0 & 1 & \cdots & 0\\
    \vdots & \vdots & \ddots & \vdots\\
    0 & 0 & \cdots & 1
    \end{pmatrix}
    = I
\end{equation}
represents the identity matrix, referred to as the stochastic resolution of the identity.\cite{takeshita2017stochastic} Here, $\mean{ \theta\otimes\theta^T}_{N_s} \equiv \frac{1}{N_s}\sum_{\xi=1}^{N_s} \theta_\xi\otimes\theta_\xi^T$ is an average over the set $\{\theta_\xi\}$ of uncorrelated stochastic orbitals $\theta_\xi$, with $\xi=1,2,\dots,N_s$. 

Analogous to the deterministic resolution of the identity (also known as density fitting), in which 3-index $(ij|A)$ and 2-index $V_{AB}=(A|B)$ ERIs are used to approximate the 4-index ERI as
\begin{equation}
    v_{ijkl} \approx \sum_{AB}^{N_\text{aux}}(ij|A)V_{AB}^{-1}(B|kl)~, 
\end{equation}
where $A(B)$ is an auxiliary basis of dimension $N_\text{aux}$, the stochastic resolution of the identity can be used as a resolution basis to approximate the 4-index ERIs as~\cite{takeshita2017stochastic}
\begin{equation}
\label{eq:sRI}
       v_{ijkl} \approx \mean{{R_{ij}} {R_{kl}}}_{N_s}~,
\end{equation}
where $R_{\alpha\beta}=\sum_{A}^{N_{aux}}(\alpha\beta|A)  \sum_{B}^{N_{aux}}V_{AB}^{-1/2}\theta_B$. One advantage of using this approximation is that the indexes $ij$ and $kl$ are decoupled, allowing to perform tensor contractions and reduce the computational scaling.\cite{takeshita2017stochastic, takeshita2019stochastic}

The use of the stochastic resolution of the identity to approximate the ERIs introduces a controllable statistical error that can be tuned by changing the number of stochastic orbitals, with a convergence rate proportional to $1/\sqrt{N_s}$. An alternative for controlling the error is to use the range-separated stochastic resolution of the identity in which the largest contributions to the ERIs are treated deterministically, while the remaining terms are treated stochastically. Specifically, as proposed in Ref. \onlinecite{dou2020range}, we first identify large contributions (denoted by the superscript $L$) to the 3-index ERIs with respect to a preset threshold,
\begin{equation}
\label{eq:epsilonp}
    (ij|A)^L = \begin{cases}
    (ij|A) & \text{if } |(ij|A)|\ge\frac{\varepsilon'}{N_{\rm e}}\{ |(ij|A)| \}^\text{max}_j \\
    0 & \text{otherwise,}
    \end{cases}
\end{equation}
where $\varepsilon'$ is a parameter in the range $[0,N_{\rm e}]$. The factor $\frac{\varepsilon'}{N}$ guarantees that the total non-vanishing elements in $(ij|A)^L$ scales as $O_{\rm e}^2)$. Then, we define the large K-tensors
\begin{equation}
    [K_{ij}^Q]^L = \sum_A^{N_\text{aux}}(ij|A)^L V_{AQ}^{-\frac{1}{2}}
\end{equation}
and keep only their larger elements, according to a second threshold
\begin{equation}
\label{eq:epsilon}
    [K_{ij}^Q]^L = \begin{cases}
    [K_{ij}^Q]^L & \text{if } |[K_{ij}^Q]^L|\ge \varepsilon\{ |[K_{ij}^Q]^L| \}^\text{max}\\
    0 & \text{otherwise,}
    \end{cases}
\end{equation}
in which $\varepsilon$ is a parameter in the range $[0,1]$. We then define large and small (denoted by the superscript $S$) R-tensors as
\begin{equation}
    R_{ij}^L = \sum_Q^{N_\text{aux}}[K_{ij}^Q]^L\theta_Q
\end{equation}
and
\begin{equation}
    R_{ij}^S = R_{ij} - R_{ij}^L~,
\end{equation}
where $R_{ij}$ is defined as in Eq.~\eqref{eq:sRI}. Using these expressions, a range-separated 4-index ERI can be written as
\begin{equation}
\label{eq:rsRI}
\begin{split}
    v_{pqrs} \approx & \sum_Q^{N_{aux}}[K_{pq}^Q]^L [K_{rs}^Q]^L + \mean{R_{pq}^L R_{rs}^S}_{N_s} + \mean{R_{pq}^S R_{rs}^L}_{N_s}\\ 
    & + \mean{R_{pq}^S R_{rs}^S}_{N_s}~.
\end{split}
\end{equation}

\subsection{Stochastic Self-Energy}
To derive a stochastic expression for the self-energy we insert Eq.~\eqref{eq:sRI} (or Eq.~\eqref{eq:rsRI} for range-separated computations) into Eqs.~\eqref{eq:selfad_final} and \eqref{eq:deltaW}, to yield
\begin{widetext}
\begin{equation}
\label{eq:self_stc}
\Sigma^{\rm ad}_{ij}[\delta\rho(t)] \approx -\frac{1}{2}\Bigg\langle\sum_{kqmn} \frac{f(\varepsilon_k)-f(\varepsilon_q)} {\varepsilon_k-\omega-\varepsilon_q-i\eta} R_{im}R_{qk} (2R'_{jn}R'_{qk}-R'_{jk}R'_{qn})\delta\rho_{mn}(t)\Bigg\rangle_{N_s},    
\end{equation}
\end{widetext}
where $\delta\rho(t)=\rho(t)-\rho(t_0)$. In the above equation, the "prime" superscript denotes that a different set of stochastic orbitals is used to construct the $R'$-tensors. Next, we rearrange the equation of motion for the density matrix (cf., Eq.~\eqref{eq:rhoEOM}) as:
\begin{equation}
\label{eq:rhoEOM_dif}
\begin{split}
i \frac{d}{dt} \rho(t)
=& \big[ F[\rho(t_0)] + \Delta H+ v^H[\delta\rho(t)] +v^{x}[\delta\rho(t)],\rho(t)\big]\\
&+ \Sigma^{\rm ad}[\delta\rho(t)]\rho(t) - \rho(t)\Sigma^{\rm ad\dag}[\delta\rho(t)]~,
\end{split}
\end{equation}
where $\Delta H=\Sigma(t_0)$ is the GF2 (or G0F2) quasiparticle energy correction. Excitation energies obtained using Eq.~\eqref{eq:rhoEOM_dif} in combination with Eq.~\eqref{eq:self_stc} will be referred to as sTD-GF2 (or sTD-G0F2).

 \begin{figure*}[t]
 \centering
\includegraphics[scale=1.0]{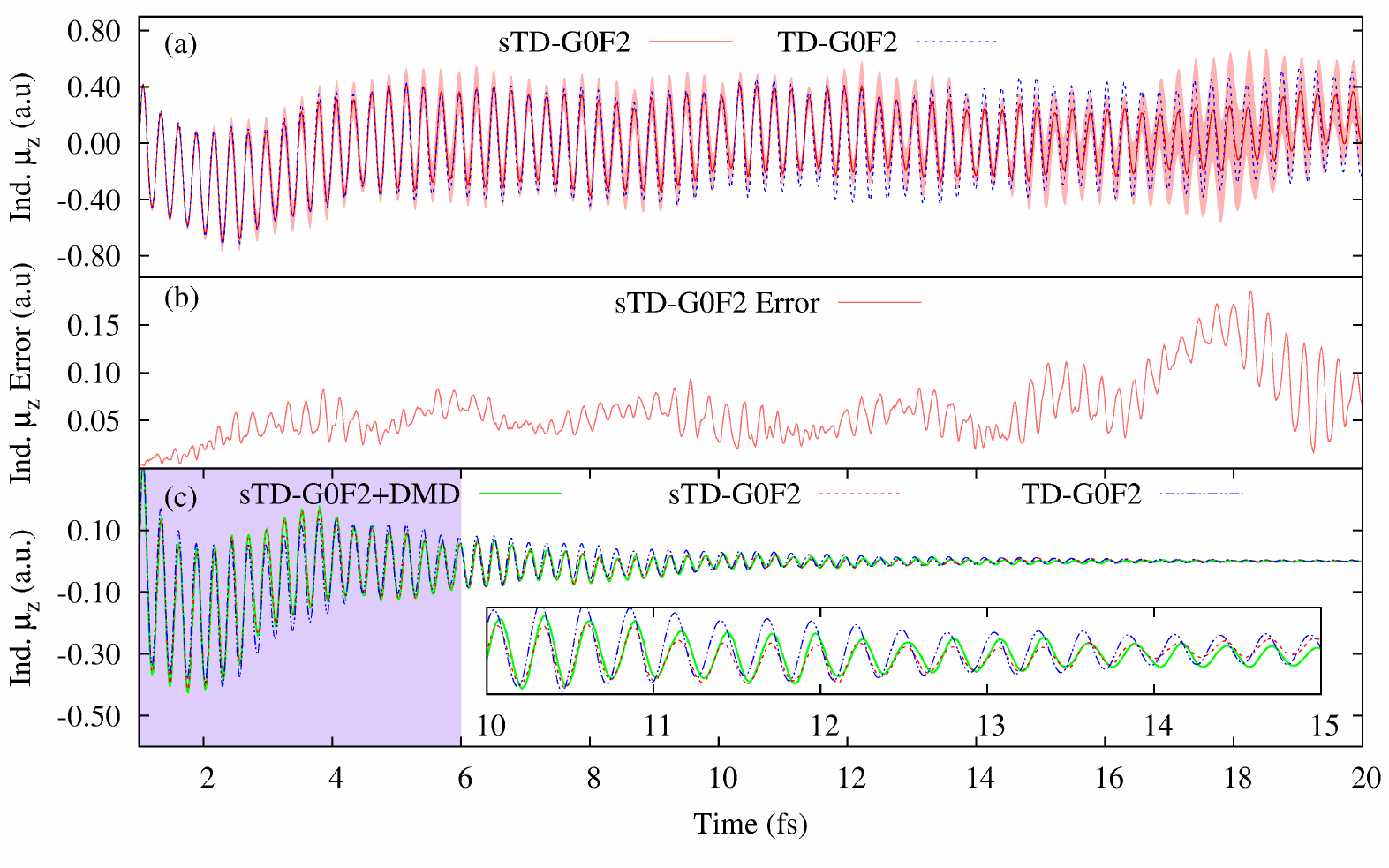}
\caption{Induced dipole dynamics and DMD extrapolation for H$_{20}$ hydrogen dimer chain using the STO-3g basis set and 80 stochastic orbitals. (a) Stochastic and deterministic time evolution of the induced dipole moment. The equation of motion was propagated using Eq.~\ref{eq:rhoEOM_dif} with stochastic (Eq.~\ref{eq:self_stc}) and deterministic (Eq.~\ref{eq:selfad_final}) self energies, respectively. The threshold parameters $\varepsilon'=20$ and $\varepsilon=1$ (fully stochastic limit) were used. The shaded red region is the standard deviation (SD) of the stochastic approach, computed from $6$ independent runs. (b) Standard error as a function of time, computed as $\text{SD}/\sqrt{N_r}$ with $N_r=6$ independent runs, for the data shown in panel (a). (c) DMD extrapolation of the stochastic induced dipole dynamics. The  shaded purple region signals the DMD window (6 fs) used for obtaining the DMD reduced model. An exponential damping function, $e^{-t/(0.1t_\text{max})}$, was added to the dynamics. Inset: zoom on the long-time dynamics.}
\label{fig:TD}
\end{figure*}

\section{\label{sec:results}Results}
To assess the accuracy of the real-time stochastic TD-GF2 formalism, we restrict the applications below to systems interacting with weak electric fields and compare the stochastic results to the linear-response GF2-BSE frequency-domain approach.\cite{doutime2022}  In the weak coupling limit, the absorption spectrum (photoabsorption cross-section) is computed by taking the imaginary part of the Fourier transform of the induced time-dependent dipole, averaged over the three spatial directions:
\begin{equation}
    \sigma(\omega) \propto \frac{1}{3}\sum_{d=x,y,z}\omega\Im\int dt e^{-i\omega t}(\text{ind. }\mu_d(t))~,
\end{equation}
where the induced dipole is given by:
\begin{equation}
    \text{ind. }\mu_d(t) = \frac{1}{\gamma}\rm{Tr}[(\rho(t)-\rho(t_0))\mu_d]~,
\end{equation}
with $d=x,y,z$ for the spatial components of the dipole moment. In the above equations, the matrix elements of the dipole operator are given by Eq.~\eqref{eq:mu_ij}, $\rho(t)$ is computed using TD-GF2 or sTD-GF2, and $\gamma\ll 1$ is a dimensionless parameter that scales the amplitude of the external electric field.

\subsection{Comparison between Deterministic and Stochastic Dynamics}
Fig.~\ref{fig:TD} shows the stochastic (sTD-G0F2) and deterministic (TD-G0F2) induced dipole dynamics of a hydrogen dimer chain (H$_{20}$, containing ten H$_{2}$ dimers with bond length of 0.74 \AA~ and intermolecular distance of 1.26 \AA, align along the $z$-axis). The equations of motion for the density matrix were propagated using Eq.~\eqref{eq:rhoEOM_dif} with stochastic (Eq.~\eqref{eq:self_stc}) and deterministic (Eq.~\eqref{eq:selfad_final}) self energies, respectively. In both cases, a Gaussian-pulse centered at $t_0=1$ fs, was used to represent the electric field, with an amplitude $\gamma E_0=0.02$ V/\AA\ and a variance of 0.005 fs; the regularization parameter appearing in Eq.~\eqref{eq:deltaW} $\eta=0.01$ and the inverse temperature is set to $\beta=50$ in all computations. For the stochastic approach, averages were computed using $N_s=80$ stochastic orbitals. In all cases, the minimal basis set STO-3G was used. 

Fig.~\ref{fig:TD}(a) exemplifies how the stochastic approach reproduces the deterministic dynamics, by comparing the TD-G0F2 and sTD-G0F2 induced dipole dynamics for the H$_{20}$ chain. The shaded region in red is the standard deviation (SD) obtained from $6$ independent runs. The statistical error can be reduced by increasing the number of stochastic orbitals (with a convergence rate proportional to $1/\sqrt{N_s}$) or by changing the range separation parameters 
$\varepsilon$ and $\varepsilon'$, as is further discussed in Sec.~\ref{sec:rs_error} below. 

We find that for a fixed number of stochastic orbitals and for fixed values of $\varepsilon$ and $\varepsilon'$, the statistical error increases with the propagation time, as shown in Fig. \ref{fig:TD}(b). This is consistent with our previous finding for the stochastic time-dependent density functional theory~\cite{neuhauser2014communication} and for the stochastic BSE approach.\cite{rabani2015time} Since the induced dipole decays rather rapidly in time, the increase in the statistical error at long times does not affect the absorption spectrum in any significant way. Nonetheless, to mitigate the divergence of the dynamics at long times, we have multiplied the induced dipole by a damping function $e^{-10t/t_\text{max}}$, where $t_\text{max}$ corresponds to the total propagation time and plays a similar role as the regularization parameter, $\eta$.

The long-time dynamics of the density matrix and the resultant time-dependent induced dipole were obtained using the DMD technique outlined above. Fig.~\ref{fig:TD}(c) shows a comparison between the extrapolated DMD dynamics and the dynamics obtained by solving Eq.~\eqref{eq:rhoEOM_dif} for both the deterministic and stochastic methods. The shaded region (first $6$ fs in Fig.~\ref{fig:TD}(c)) indicates the portion of the dynamics that was used to train the reduced DMD model (defined as \emph{DMD window}), while the remaining $34$ fs (of which 14 fs are shown in Fig.~\ref{fig:TD}(c)) corresponds to the extrapolated dynamics. We find that the DMD technique accurately captures the main dynamical features, even for the noisy stochastic data. Naturally, the time scale of the events that can be captured by the reduced DMD model depends on the DMD window length. Below, we analyze the accuracy of the DMD approach in reproducing the absorption spectra.

 \begin{figure*}[t]
 \centering
\includegraphics[scale=1.0]{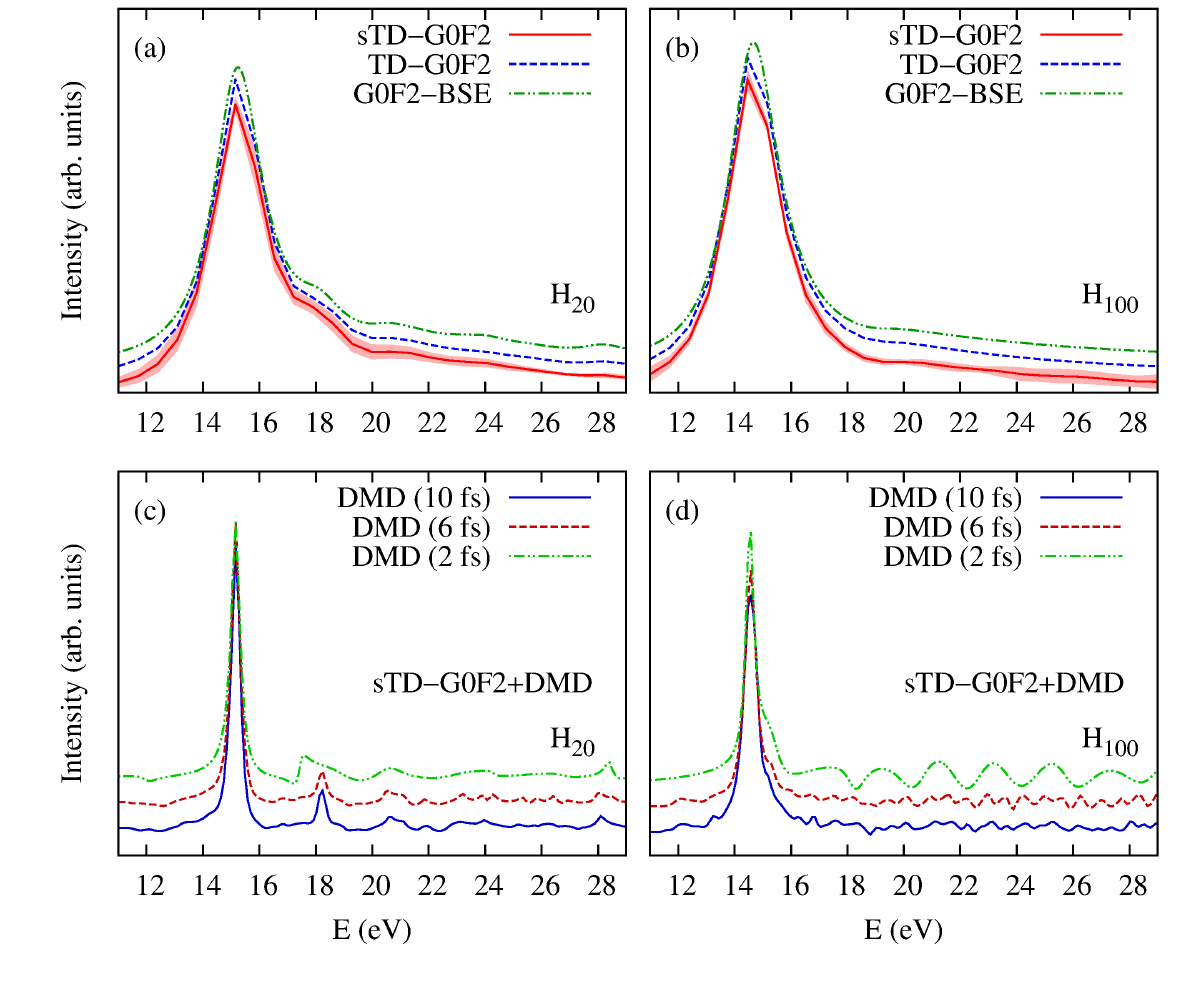}
\caption{Absorption spectra for two representative Hydrogen dimer chains with varying lengths. (a)-(b) Computed from 6 fs real-time stochastic and deterministic dynamics, and their comparison with the linear-response equivalent in the frequency domain (G0F2-BSE). (c)-(d) Computed from stochastic trajectories with DMD extrapolation for varying DMD window lengths. In all cases, the absorption spectra were shifted vertically for clarity, $N_s=80$, $\varepsilon'=20$ and 100 for H$_{20}$ and H$_{100}$, respectively, $\varepsilon=1$ (fully stochastic limit), and STO-3G was used as the basis set.}
\label{fig:spectra}
\end{figure*}

\subsection{Comparison between Deterministic and Stochastic Absorption Spectra}

In Fig.~\ref{fig:spectra} panels (a) and (b) we compare the absorption spectra obtained from the stochastic and deterministic real-time dynamics and the reference deterministic frequency-domain linear-response approach (G0F2-BSE), for two representative hydrogen dimer chains. The absorption spectra obtained from the three different approaches (vertically shifted for clarity) are numerically identical, demonstrating that the real-time implementations are consistent with the frequency domain reference methods (in the weak coupling-linear response limit) and, in particular, that the stochastic approach can reproduce the benchmark results with only $80$ stochastic orbitals.

The frequency resolution of the absorption spectra can be improved by propagating the density matrix dynamics to longer times using the DMD technique. Fig.~\ref{fig:spectra} panels (c) and (d) show the corresponding absorption spectra for a $40$ fs extrapolated trajectory (sTD-G0F2+DMD) considering three different DMD window lengths. Even a short ($2$fs) window length provides a reasonable description of the low-excitation features (main absorption peak at $\sim$15 eV), but the quality of the spectra improves with increasing DMD windows lengths, especially for the higher-excitation peaks. Specifically, for the sTD-G0F2+DMD method, we observed that the average DMD spectral error is proportional to $1/\sqrt{t_\text{DMD}}$, with $t_\text{DMD}$ being the DMD window length. 

\subsection{Error analysis and scaling}
\label{sec:rs_error}
 \begin{figure}[t]
 \centering
\includegraphics[scale=1.0]{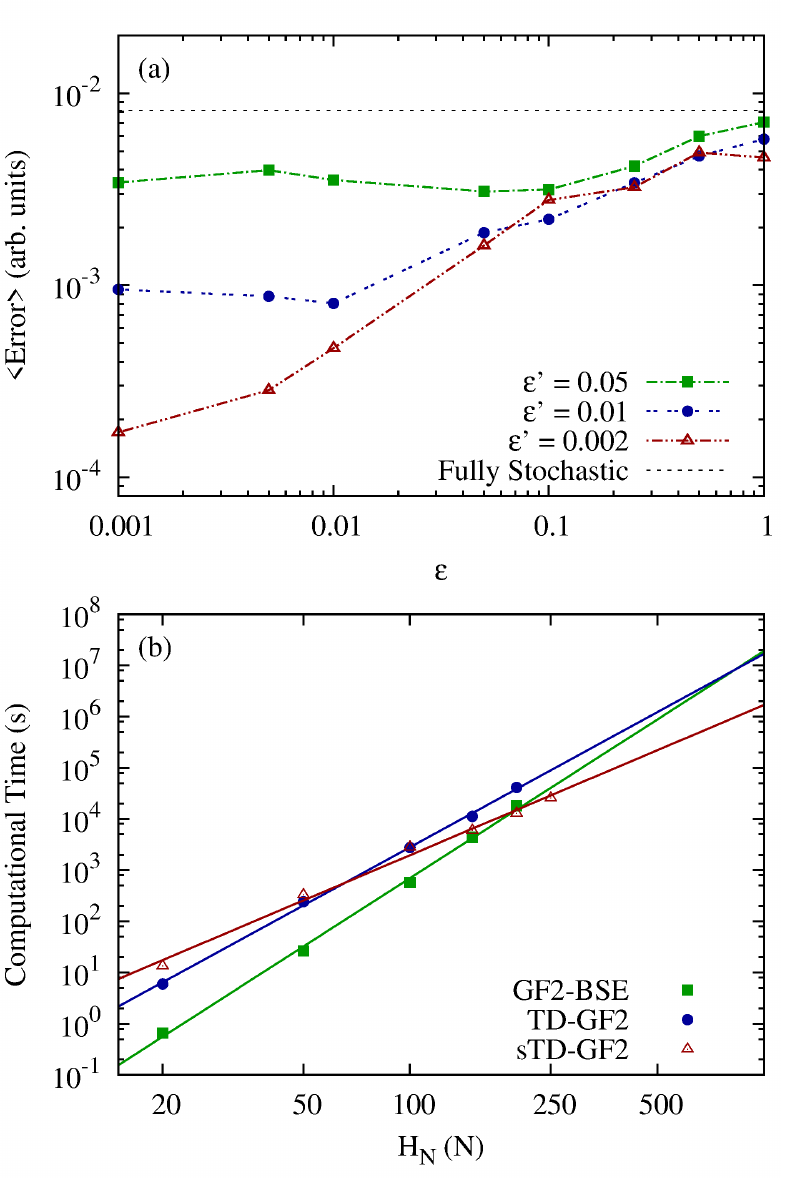}
\caption{(a) The average spectrum error for the H$_{20}$ hydrogen dimer chain using the range-separated sTD-G0F2 method for varying threshold parameters, $\varepsilon'$ and $\varepsilon$. (b) Log-log plot of the computational cost as a function of system size for hydrogen dimer chains with varying lengths. For time-dependent methods, the propagation time was 2 fs. For the stochastic computations, threshold parameters $\varepsilon'=N$ and $\varepsilon=1$ (stochastic limit) were used. The observed scaling with system size is $O(N_{\rm e}^3)$ for sTD-GF2, $O(N_{\rm e}^4)$ for TD-GF2, and $O(N_{\rm e}^{4.5})$ for GF2-BSE. In all cases, $N_s=80$ and the STO-3G basis set were used.}
\label{fig:rs}
\end{figure}

The variation of the range-separated threshold parameters, $\varepsilon$ and $\varepsilon'$ (see Eqs.~\eqref{eq:epsilonp} and \eqref{eq:epsilon}), allows us to control the ratio of deterministic to stochastic Coulomb tensor elements. As $\varepsilon' \to N$ or $\varepsilon\to 1$ the approach reduces to the fully stochastic limit. By contrast, when $\varepsilon'\to 0$ or $\varepsilon\to 0$ the approach is fully deterministic. Fig.~\ref{fig:rs}(a) shows the dependence of the statistical error on $\varepsilon'$ and $\varepsilon$ for H$_{20}$. The average error was estimated using $n=6$ independent stochastic runs as
\begin{equation}
    \mean{Error} = \frac{1}{N_\omega}\sum_\omega^{N_\omega}\frac{1}{n}\sqrt{\sum_i^n(\sigma_i(\omega)-\mean{\sigma(\omega)})^2}~,
\end{equation}
where $N_\omega$ is the number of frequencies used in the range $E=10-30$ eV. As $\varepsilon$ increases the statistical error increases and approaches the fully stochastic limit (dotted line in Fig.~\ref{fig:rs}(a)). For the case with the lowest statistical error in Fig. \ref{fig:rs}(a) ($\varepsilon'=0.002$, $\varepsilon=0.001$), the amount of ERI elements computed deterministically corresponds to $\approx 10\%$ for H$_{20}$, resulting in an error reduction of almost $2$ orders of magnitude compared to the fully stochastic limit.

The main advantage of using the stochastic formulation of GF2 in the real-time domain is the reduction in the computational complexity and scaling. Formally, GF2 in the frequency-domain scales as $O(N_{\rm e}^6)$ with the system size ($N$) while the real-time deterministic implementation scales as $O_{\rm e}^5)$. By contrast, when the stochastic resolution of identity is used in the time-domain, the computational scaling is further reduced to $O_{\rm e}^3)$, as long as the number of stochastic orbitals does not increase with system size to achieve a similar statistical error (which is the case for the systems studied here). The computational limiting step in the sTD-GF2 method is the computation of the self-energy (Eq.~\eqref{eq:self_stc}), with a formal scaling of $O(N_s N_{\rm e}^3)$ when appropriate tensor contractions are used. Figure \ref{fig:rs}(b) shows the computational cost associated with the stochastic and deterministic real-time methods and the equivalent frequency-domain linear-response implementation for hydrogen dimer chains with varying lengths. The lowest scaling corresponds to the stochastic real-time implementation, sTD-GF2, which exhibits an $O_{\rm e}^3)$ behavior, with a large pre-factor. For the current target statistical error, the stochastic approach is computationally more efficient than the deterministic approach for system sizes that exceed $N \approx 200$ basis functions.

\section{\label{sec:conclusions}Conclusions}
We presented a stochastic real-time approach to compute excited state energies in extended systems based on the adiabatic approximation to the  Kadanoff-Baym equations using the second-order Born approximation to the self-energy (referred to as sTD-GF2). We showed that the sTD-GF2 approach reproduces the benchmark linear-response results from analogous deterministic methods, namely TD-GF2 and GF2-BSE,\cite{doutime2022} but at a much milder computational cost that scales as $O(N_{\rm e}^3)$ with system size, in contrast to the formal $O(N_{\rm e}^5)$ and $O(N_{\rm e}^6)$ of TD-GF2 and GF2-BSE, respectively. The reduction in scaling is achieved by introducing a statistical error that can be controlled by varying the number of stochastic orbitals or by tuning the fraction of ERIs that are computed deterministically using the range-separated resolution of the identity.

Within the adiabatic approximation, the Kadanoff-Baym equations can be reduced to a single-time integro-differential equation, which is efficiently solved using the dynamic mode decomposition method. We assessed the performance of the DMD method for a chain of hydrogen dimers of various lengths and found that it is sufficient to train the systems for times as short as 2 fs (independent of the system size) to greatly improve the resolution of the absorption spectra.

The method presented in this work offers the possibility to study neutral excitations in systems with hundreds to thousands of electrons at the GF2 closure. This complements the growing manifold of stochastic methods capable of elucidating the electronic structure of the ground and excited states in extended systems with open or closed boundary conditions. Further directions include the development of stochastic techniques that allow the efficient propagation of the two-time Kadanoff-Baym equations (Eq.~\eqref{eq:kb1} and \ref{eq:kb2}), opening the possibility to describe strongly driven system beyond the adiabatic limit.

\acknowledgements
CY, DRR, and ER are grateful for support from the U.S. Department of Energy, Office of Science, Office of Advanced Scientific Computing Research, Scientific Discovery through Advanced Computing (SciDAC) program, under Award No. DE-SC0022088. Some of the methods used in this work were provided by the Center for Computational Study of Excited State Phenomena in Energy Materials (C2SEPEM), which is funded
by the U.S. Department of Energy, Office of Science, Basic Energy Sciences, Materials Sciences and Engineering Division, via Contract No. DE-AC02-05CH11231 as part of the Computational Materials Sciences program. Resources of the National Energy Research Scientific Computing Center (NERSC), a U.S. Department of Energy Office of Science User Facility operated under Contract No. DE-AC02-05CH11231 are also acknowledged. RB and ER acknowledge support from the US-Israel Binational Science Foundation BSF-201836.

\bibliography{references}

\providecommand{\noopsort}[1]{}\providecommand{\singleletter}[1]{#1}%
\begin{thebibliography}{49}%
\makeatletter
\providecommand \@ifxundefined [1]{%
 \@ifx{#1\undefined}
}%
\providecommand \@ifnum [1]{%
 \ifnum #1\expandafter \@firstoftwo
 \else \expandafter \@secondoftwo
 \fi
}%
\providecommand \@ifx [1]{%
 \ifx #1\expandafter \@firstoftwo
 \else \expandafter \@secondoftwo
 \fi
}%
\providecommand \natexlab [1]{#1}%
\providecommand \enquote  [1]{``#1''}%
\providecommand \bibnamefont  [1]{#1}%
\providecommand \bibfnamefont [1]{#1}%
\providecommand \citenamefont [1]{#1}%
\providecommand \href@noop [0]{\@secondoftwo}%
\providecommand \href [0]{\begingroup \@sanitize@url \@href}%
\providecommand \@href[1]{\@@startlink{#1}\@@href}%
\providecommand \@@href[1]{\endgroup#1\@@endlink}%
\providecommand \@sanitize@url [0]{\catcode `\\12\catcode `\$12\catcode
  `\&12\catcode `\#12\catcode `\^12\catcode `\_12\catcode `\%12\relax}%
\providecommand \@@startlink[1]{}%
\providecommand \@@endlink[0]{}%
\providecommand \url  [0]{\begingroup\@sanitize@url \@url }%
\providecommand \@url [1]{\endgroup\@href {#1}{\urlprefix }}%
\providecommand \urlprefix  [0]{URL }%
\providecommand \Eprint [0]{\href }%
\providecommand \doibase [0]{https://doi.org/}%
\providecommand \selectlanguage [0]{\@gobble}%
\providecommand \bibinfo  [0]{\@secondoftwo}%
\providecommand \bibfield  [0]{\@secondoftwo}%
\providecommand \translation [1]{[#1]}%
\providecommand \BibitemOpen [0]{}%
\providecommand \bibitemStop [0]{}%
\providecommand \bibitemNoStop [0]{.\EOS\space}%
\providecommand \EOS [0]{\spacefactor3000\relax}%
\providecommand \BibitemShut  [1]{\csname bibitem#1\endcsname}%
\let\auto@bib@innerbib\@empty
\bibitem [{\citenamefont {Adamo}\ and\ \citenamefont
  {Jacquemin}(2013)}]{adamo2013calculations}%
  \BibitemOpen
  \bibfield  {author} {\bibinfo {author} {\bibfnamefont {C.}~\bibnamefont
  {Adamo}}\ and\ \bibinfo {author} {\bibfnamefont {D.}~\bibnamefont
  {Jacquemin}},\ }\bibfield  {title} {\enquote {\bibinfo {title} {The
  calculations of excited-state properties with time-dependent density
  functional theory},}\ }\href@noop {} {\bibfield  {journal} {\bibinfo
  {journal} {Chem. Soc. Rev.}\ }\textbf {\bibinfo {volume} {42}},\ \bibinfo
  {pages} {845--856} (\bibinfo {year} {2013})}\BibitemShut {NoStop}%
\bibitem [{\citenamefont {Gonz{\'a}lez}, \citenamefont {Escudero},\ and\
  \citenamefont {Serrano-Andr{\'e}s}(2012)}]{gonzalez2012progress}%
  \BibitemOpen
  \bibfield  {author} {\bibinfo {author} {\bibfnamefont {L.}~\bibnamefont
  {Gonz{\'a}lez}}, \bibinfo {author} {\bibfnamefont {D.}~\bibnamefont
  {Escudero}},\ and\ \bibinfo {author} {\bibfnamefont {L.}~\bibnamefont
  {Serrano-Andr{\'e}s}},\ }\bibfield  {title} {\enquote {\bibinfo {title}
  {Progress and challenges in the calculation of electronic excited states},}\
  }\href@noop {} {\bibfield  {journal} {\bibinfo  {journal} {ChemPhysChem}\
  }\textbf {\bibinfo {volume} {13}},\ \bibinfo {pages} {28--51} (\bibinfo
  {year} {2012})}\BibitemShut {NoStop}%
\bibitem [{\citenamefont {Lischka}\ \emph {et~al.}(2018)\citenamefont
  {Lischka}, \citenamefont {Nachtigallova}, \citenamefont {Aquino},
  \citenamefont {Szalay}, \citenamefont {Plasser}, \citenamefont {Machado},\
  and\ \citenamefont {Barbatti}}]{lischka2018multireference}%
  \BibitemOpen
  \bibfield  {author} {\bibinfo {author} {\bibfnamefont {H.}~\bibnamefont
  {Lischka}}, \bibinfo {author} {\bibfnamefont {D.}~\bibnamefont
  {Nachtigallova}}, \bibinfo {author} {\bibfnamefont {A.~J.}\ \bibnamefont
  {Aquino}}, \bibinfo {author} {\bibfnamefont {P.~G.}\ \bibnamefont {Szalay}},
  \bibinfo {author} {\bibfnamefont {F.}~\bibnamefont {Plasser}}, \bibinfo
  {author} {\bibfnamefont {F.~B.}\ \bibnamefont {Machado}},\ and\ \bibinfo
  {author} {\bibfnamefont {M.}~\bibnamefont {Barbatti}},\ }\bibfield  {title}
  {\enquote {\bibinfo {title} {Multireference approaches for excited states of
  molecules},}\ }\href@noop {} {\bibfield  {journal} {\bibinfo  {journal}
  {Chem. Rev.}\ }\textbf {\bibinfo {volume} {118}},\ \bibinfo {pages}
  {7293--7361} (\bibinfo {year} {2018})}\BibitemShut {NoStop}%
\bibitem [{\citenamefont {Dreuw}\ and\ \citenamefont
  {Wormit}(2015)}]{dreuw2015algebraic}%
  \BibitemOpen
  \bibfield  {author} {\bibinfo {author} {\bibfnamefont {A.}~\bibnamefont
  {Dreuw}}\ and\ \bibinfo {author} {\bibfnamefont {M.}~\bibnamefont {Wormit}},\
  }\bibfield  {title} {\enquote {\bibinfo {title} {The algebraic diagrammatic
  construction scheme for the polarization propagator for the calculation of
  excited states},}\ }\href@noop {} {\bibfield  {journal} {\bibinfo  {journal}
  {Wiley Interdiscip. Rev. Comput. Mol. Sci.}\ }\textbf {\bibinfo {volume}
  {5}},\ \bibinfo {pages} {82--95} (\bibinfo {year} {2015})}\BibitemShut
  {NoStop}%
\bibitem [{\citenamefont {Hernandez}\ \emph {et~al.}(2018)\citenamefont
  {Hernandez}, \citenamefont {Xia}, \citenamefont {Vl{\v{c}}ek}, \citenamefont
  {Boutelle}, \citenamefont {Baer}, \citenamefont {Rabani},\ and\ \citenamefont
  {Neuhauser}}]{hernandez2018first}%
  \BibitemOpen
  \bibfield  {author} {\bibinfo {author} {\bibfnamefont {S.}~\bibnamefont
  {Hernandez}}, \bibinfo {author} {\bibfnamefont {Y.}~\bibnamefont {Xia}},
  \bibinfo {author} {\bibfnamefont {V.}~\bibnamefont {Vl{\v{c}}ek}}, \bibinfo
  {author} {\bibfnamefont {R.}~\bibnamefont {Boutelle}}, \bibinfo {author}
  {\bibfnamefont {R.}~\bibnamefont {Baer}}, \bibinfo {author} {\bibfnamefont
  {E.}~\bibnamefont {Rabani}},\ and\ \bibinfo {author} {\bibfnamefont
  {D.}~\bibnamefont {Neuhauser}},\ }\bibfield  {title} {\enquote {\bibinfo
  {title} {First-principles spectra of au nanoparticles: from quantum to
  classical absorption},}\ }\href@noop {} {\bibfield  {journal} {\bibinfo
  {journal} {Mol. Phys.}\ }\textbf {\bibinfo {volume} {116}},\ \bibinfo {pages}
  {2506--2511} (\bibinfo {year} {2018})}\BibitemShut {NoStop}%
\bibitem [{\citenamefont {Dou}\ \emph {et~al.}(2022)\citenamefont {Dou},
  \citenamefont {Lee}, \citenamefont {Zhu}, \citenamefont {Mejía},
  \citenamefont {Reichman}, \citenamefont {Baer},\ and\ \citenamefont
  {Rabani}}]{doutime2022}%
  \BibitemOpen
  \bibfield  {author} {\bibinfo {author} {\bibfnamefont {W.}~\bibnamefont
  {Dou}}, \bibinfo {author} {\bibfnamefont {J.}~\bibnamefont {Lee}}, \bibinfo
  {author} {\bibfnamefont {J.}~\bibnamefont {Zhu}}, \bibinfo {author}
  {\bibfnamefont {L.}~\bibnamefont {Mejía}}, \bibinfo {author} {\bibfnamefont
  {D.~R.}\ \bibnamefont {Reichman}}, \bibinfo {author} {\bibfnamefont
  {R.}~\bibnamefont {Baer}},\ and\ \bibinfo {author} {\bibfnamefont
  {E.}~\bibnamefont {Rabani}},\ }\bibfield  {title} {\enquote {\bibinfo {title}
  {Time-dependent second-order green’s function theory for neutral
  excitations},}\ }\href {https://doi.org/10.1021/acs.jctc.2c00057} {\bibfield
  {journal} {\bibinfo  {journal} {J. Chem. Theory Comput.}\ }\textbf {\bibinfo
  {volume} {18}},\ \bibinfo {pages} {5221--5232} (\bibinfo {year} {2022})},\
  \bibinfo {note} {pMID: 36040050},\ \Eprint
  {https://arxiv.org/abs/https://doi.org/10.1021/acs.jctc.2c00057}
  {https://doi.org/10.1021/acs.jctc.2c00057} \BibitemShut {NoStop}%
\bibitem [{\citenamefont {Jasrasaria}\ \emph {et~al.}(2020)\citenamefont
  {Jasrasaria}, \citenamefont {Philbin}, \citenamefont {Yan}, \citenamefont
  {Weinberg}, \citenamefont {Alivisatos},\ and\ \citenamefont
  {Rabani}}]{jasrasaria2020sub}%
  \BibitemOpen
  \bibfield  {author} {\bibinfo {author} {\bibfnamefont {D.}~\bibnamefont
  {Jasrasaria}}, \bibinfo {author} {\bibfnamefont {J.~P.}\ \bibnamefont
  {Philbin}}, \bibinfo {author} {\bibfnamefont {C.}~\bibnamefont {Yan}},
  \bibinfo {author} {\bibfnamefont {D.}~\bibnamefont {Weinberg}}, \bibinfo
  {author} {\bibfnamefont {A.~P.}\ \bibnamefont {Alivisatos}},\ and\ \bibinfo
  {author} {\bibfnamefont {E.}~\bibnamefont {Rabani}},\ }\bibfield  {title}
  {\enquote {\bibinfo {title} {Sub-bandgap photoinduced transient absorption
  features in cdse nanostructures: The role of trapped holes},}\ }\href@noop {}
  {\bibfield  {journal} {\bibinfo  {journal} {J. Phys. Chem. C}\ }\textbf
  {\bibinfo {volume} {124}},\ \bibinfo {pages} {17372--17378} (\bibinfo {year}
  {2020})}\BibitemShut {NoStop}%
\bibitem [{\citenamefont {Del~Ben}\ \emph {et~al.}(2020)\citenamefont
  {Del~Ben}, \citenamefont {Yang}, \citenamefont {Li}, \citenamefont {Felipe},
  \citenamefont {Louie},\ and\ \citenamefont {Deslippe}}]{del2020accelerating}%
  \BibitemOpen
  \bibfield  {author} {\bibinfo {author} {\bibfnamefont {M.}~\bibnamefont
  {Del~Ben}}, \bibinfo {author} {\bibfnamefont {C.}~\bibnamefont {Yang}},
  \bibinfo {author} {\bibfnamefont {Z.}~\bibnamefont {Li}}, \bibinfo {author}
  {\bibfnamefont {H.}~\bibnamefont {Felipe}}, \bibinfo {author} {\bibfnamefont
  {S.~G.}\ \bibnamefont {Louie}},\ and\ \bibinfo {author} {\bibfnamefont
  {J.}~\bibnamefont {Deslippe}},\ }\bibfield  {title} {\enquote {\bibinfo
  {title} {Accelerating large-scale excited-state gw calculations on leadership
  hpc systems},}\ }in\ \href@noop {} {\emph {\bibinfo {booktitle} {SC20:
  International Conference for High Performance Computing, Networking, Storage
  and Analysis}}}\ (\bibinfo {organization} {IEEE},\ \bibinfo {year} {2020})\
  pp.\ \bibinfo {pages} {1--11}\BibitemShut {NoStop}%
\bibitem [{\citenamefont {Liang}\ \emph {et~al.}(2022)\citenamefont {Liang},
  \citenamefont {Feng}, \citenamefont {Hait},\ and\ \citenamefont
  {Head-Gordon}}]{liang2022revisiting}%
  \BibitemOpen
  \bibfield  {author} {\bibinfo {author} {\bibfnamefont {J.}~\bibnamefont
  {Liang}}, \bibinfo {author} {\bibfnamefont {X.}~\bibnamefont {Feng}},
  \bibinfo {author} {\bibfnamefont {D.}~\bibnamefont {Hait}},\ and\ \bibinfo
  {author} {\bibfnamefont {M.}~\bibnamefont {Head-Gordon}},\ }\bibfield
  {title} {\enquote {\bibinfo {title} {Revisiting the performance of
  time-dependent density functional theory for electronic excitations:
  Assessment of 43 popular and recently developed functionals from rungs one to
  four},}\ }\href@noop {} {\bibfield  {journal} {\bibinfo  {journal} {J. Chem.
  Theory Comput.}\ }\textbf {\bibinfo {volume} {18}},\ \bibinfo {pages}
  {3460--3473} (\bibinfo {year} {2022})}\BibitemShut {NoStop}%
\bibitem [{\citenamefont {Hait}\ and\ \citenamefont
  {Head-Gordon}(2021)}]{hait2021orbital}%
  \BibitemOpen
  \bibfield  {author} {\bibinfo {author} {\bibfnamefont {D.}~\bibnamefont
  {Hait}}\ and\ \bibinfo {author} {\bibfnamefont {M.}~\bibnamefont
  {Head-Gordon}},\ }\bibfield  {title} {\enquote {\bibinfo {title} {Orbital
  optimized density functional theory for electronic excited states},}\
  }\href@noop {} {\bibfield  {journal} {\bibinfo  {journal} {J. Phys. Chem.
  Lett.}\ }\textbf {\bibinfo {volume} {12}},\ \bibinfo {pages} {4517--4529}
  (\bibinfo {year} {2021})}\BibitemShut {NoStop}%
\bibitem [{\citenamefont {Higgott}, \citenamefont {Wang},\ and\ \citenamefont
  {Brierley}(2019)}]{higgott2019variational}%
  \BibitemOpen
  \bibfield  {author} {\bibinfo {author} {\bibfnamefont {O.}~\bibnamefont
  {Higgott}}, \bibinfo {author} {\bibfnamefont {D.}~\bibnamefont {Wang}},\ and\
  \bibinfo {author} {\bibfnamefont {S.}~\bibnamefont {Brierley}},\ }\bibfield
  {title} {\enquote {\bibinfo {title} {Variational quantum computation of
  excited states},}\ }\href@noop {} {\bibfield  {journal} {\bibinfo  {journal}
  {Quantum}\ }\textbf {\bibinfo {volume} {3}},\ \bibinfo {pages} {156}
  (\bibinfo {year} {2019})}\BibitemShut {NoStop}%
\bibitem [{\citenamefont {Faber}\ \emph {et~al.}(2014)\citenamefont {Faber},
  \citenamefont {Boulanger}, \citenamefont {Attaccalite}, \citenamefont
  {Duchemin},\ and\ \citenamefont {Blase}}]{faber2014excited}%
  \BibitemOpen
  \bibfield  {author} {\bibinfo {author} {\bibfnamefont {C.}~\bibnamefont
  {Faber}}, \bibinfo {author} {\bibfnamefont {P.}~\bibnamefont {Boulanger}},
  \bibinfo {author} {\bibfnamefont {C.}~\bibnamefont {Attaccalite}}, \bibinfo
  {author} {\bibfnamefont {I.}~\bibnamefont {Duchemin}},\ and\ \bibinfo
  {author} {\bibfnamefont {X.}~\bibnamefont {Blase}},\ }\bibfield  {title}
  {\enquote {\bibinfo {title} {Excited states properties of organic molecules:
  From density functional theory to the gw and bethe--salpeter green's function
  formalisms},}\ }\href@noop {} {\bibfield  {journal} {\bibinfo  {journal}
  {Philos. Trans. R. Soc. A}\ }\textbf {\bibinfo {volume} {372}},\ \bibinfo
  {pages} {20130271} (\bibinfo {year} {2014})}\BibitemShut {NoStop}%
\bibitem [{\citenamefont {Marques}\ and\ \citenamefont
  {Gross}(2004)}]{marques2004time}%
  \BibitemOpen
  \bibfield  {author} {\bibinfo {author} {\bibfnamefont {M.~A.}\ \bibnamefont
  {Marques}}\ and\ \bibinfo {author} {\bibfnamefont {E.~K.}\ \bibnamefont
  {Gross}},\ }\bibfield  {title} {\enquote {\bibinfo {title} {Time-dependent
  density functional theory},}\ }\href@noop {} {\bibfield  {journal} {\bibinfo
  {journal} {Annu. Rev. Phys. Chem.}\ }\textbf {\bibinfo {volume} {55}},\
  \bibinfo {pages} {427--455} (\bibinfo {year} {2004})}\BibitemShut {NoStop}%
\bibitem [{\citenamefont {Burke}, \citenamefont {Werschnik},\ and\
  \citenamefont {Gross}(2005)}]{burke2005time}%
  \BibitemOpen
  \bibfield  {author} {\bibinfo {author} {\bibfnamefont {K.}~\bibnamefont
  {Burke}}, \bibinfo {author} {\bibfnamefont {J.}~\bibnamefont {Werschnik}},\
  and\ \bibinfo {author} {\bibfnamefont {E.}~\bibnamefont {Gross}},\ }\bibfield
   {title} {\enquote {\bibinfo {title} {Time-dependent density functional
  theory: Past, present, and future},}\ }\href@noop {} {\bibfield  {journal}
  {\bibinfo  {journal} {J. Chem. Phys.}\ }\textbf {\bibinfo {volume} {123}},\
  \bibinfo {pages} {062206} (\bibinfo {year} {2005})}\BibitemShut {NoStop}%
\bibitem [{\citenamefont {Casida}\ and\ \citenamefont
  {Huix-Rotllant}(2012)}]{casida2012progress}%
  \BibitemOpen
  \bibfield  {author} {\bibinfo {author} {\bibfnamefont {M.~E.}\ \bibnamefont
  {Casida}}\ and\ \bibinfo {author} {\bibfnamefont {M.}~\bibnamefont
  {Huix-Rotllant}},\ }\bibfield  {title} {\enquote {\bibinfo {title} {Progress
  in time-dependent density-functional theory},}\ }\href@noop {} {\bibfield
  {journal} {\bibinfo  {journal} {Annu. Rev. Phys. Chem.}\ }\textbf {\bibinfo
  {volume} {63}},\ \bibinfo {pages} {287--323} (\bibinfo {year}
  {2012})}\BibitemShut {NoStop}%
\bibitem [{\citenamefont {Marques}\ \emph {et~al.}(2006)\citenamefont
  {Marques}, \citenamefont {Ullrich}, \citenamefont {Nogueira}, \citenamefont
  {Rubio}, \citenamefont {Burke},\ and\ \citenamefont
  {Gross}}]{marques2006time2}%
  \BibitemOpen
  \bibfield  {author} {\bibinfo {author} {\bibfnamefont {M.~A.}\ \bibnamefont
  {Marques}}, \bibinfo {author} {\bibfnamefont {C.~A.}\ \bibnamefont
  {Ullrich}}, \bibinfo {author} {\bibfnamefont {F.}~\bibnamefont {Nogueira}},
  \bibinfo {author} {\bibfnamefont {A.}~\bibnamefont {Rubio}}, \bibinfo
  {author} {\bibfnamefont {K.}~\bibnamefont {Burke}},\ and\ \bibinfo {author}
  {\bibfnamefont {E.~K.}\ \bibnamefont {Gross}},\ }\href@noop {} {\emph
  {\bibinfo {title} {Time-Dependent Density Functional Theory}}},\ Vol.\
  \bibinfo {volume} {706}\ (\bibinfo  {publisher} {Springer Science \& Business
  Media},\ \bibinfo {year} {2006})\BibitemShut {NoStop}%
\bibitem [{\citenamefont {McLachlan}\ and\ \citenamefont
  {Ball}(1964)}]{mclachlan1964time}%
  \BibitemOpen
  \bibfield  {author} {\bibinfo {author} {\bibfnamefont {A.}~\bibnamefont
  {McLachlan}}\ and\ \bibinfo {author} {\bibfnamefont {M.}~\bibnamefont
  {Ball}},\ }\bibfield  {title} {\enquote {\bibinfo {title} {Time-dependent
  hartree—fock theory for molecules},}\ }\href@noop {} {\bibfield  {journal}
  {\bibinfo  {journal} {Rev. Mod. Phys.}\ }\textbf {\bibinfo {volume} {36}},\
  \bibinfo {pages} {844} (\bibinfo {year} {1964})}\BibitemShut {NoStop}%
\bibitem [{\citenamefont {Jorgensen}(1975)}]{jorgensen1975molecular}%
  \BibitemOpen
  \bibfield  {author} {\bibinfo {author} {\bibfnamefont {P.}~\bibnamefont
  {Jorgensen}},\ }\bibfield  {title} {\enquote {\bibinfo {title} {Molecular and
  atomic applications of time-dependent hartree-fock theory},}\ }\href@noop {}
  {\bibfield  {journal} {\bibinfo  {journal} {Annu. Rev. Phys. Chem.}\ }\textbf
  {\bibinfo {volume} {26}},\ \bibinfo {pages} {359--380} (\bibinfo {year}
  {1975})}\BibitemShut {NoStop}%
\bibitem [{\citenamefont {Li}\ \emph {et~al.}(2005)\citenamefont {Li},
  \citenamefont {Smith}, \citenamefont {Markevitch}, \citenamefont {Romanov},
  \citenamefont {Levis},\ and\ \citenamefont {Schlegel}}]{li2005time}%
  \BibitemOpen
  \bibfield  {author} {\bibinfo {author} {\bibfnamefont {X.}~\bibnamefont
  {Li}}, \bibinfo {author} {\bibfnamefont {S.~M.}\ \bibnamefont {Smith}},
  \bibinfo {author} {\bibfnamefont {A.~N.}\ \bibnamefont {Markevitch}},
  \bibinfo {author} {\bibfnamefont {D.~A.}\ \bibnamefont {Romanov}}, \bibinfo
  {author} {\bibfnamefont {R.~J.}\ \bibnamefont {Levis}},\ and\ \bibinfo
  {author} {\bibfnamefont {H.~B.}\ \bibnamefont {Schlegel}},\ }\bibfield
  {title} {\enquote {\bibinfo {title} {A time-dependent hartree--fock approach
  for studying the electronic optical response of molecules in intense
  fields},}\ }\href@noop {} {\bibfield  {journal} {\bibinfo  {journal} {Phys.
  Chem. Chem. Phys.}\ }\textbf {\bibinfo {volume} {7}},\ \bibinfo {pages}
  {233--239} (\bibinfo {year} {2005})}\BibitemShut {NoStop}%
\bibitem [{\citenamefont {Hirata}(2004)}]{hirata2004higher}%
  \BibitemOpen
  \bibfield  {author} {\bibinfo {author} {\bibfnamefont {S.}~\bibnamefont
  {Hirata}},\ }\bibfield  {title} {\enquote {\bibinfo {title} {Higher-order
  equation-of-motion coupled-cluster methods},}\ }\href@noop {} {\bibfield
  {journal} {\bibinfo  {journal} {J. Chem. Phys.}\ }\textbf {\bibinfo {volume}
  {121}},\ \bibinfo {pages} {51--59} (\bibinfo {year} {2004})}\BibitemShut
  {NoStop}%
\bibitem [{\citenamefont {Bartlett}(2012)}]{bartlett2012coupled}%
  \BibitemOpen
  \bibfield  {author} {\bibinfo {author} {\bibfnamefont {R.~J.}\ \bibnamefont
  {Bartlett}},\ }\bibfield  {title} {\enquote {\bibinfo {title}
  {Coupled-cluster theory and its equation-of-motion extensions},}\ }\href@noop
  {} {\bibfield  {journal} {\bibinfo  {journal} {Wiley Interdiscip. Rev.
  Comput. Mol. Sci.}\ }\textbf {\bibinfo {volume} {2}},\ \bibinfo {pages}
  {126--138} (\bibinfo {year} {2012})}\BibitemShut {NoStop}%
\bibitem [{\citenamefont {Krylov}(2008)}]{krylov2008equation}%
  \BibitemOpen
  \bibfield  {author} {\bibinfo {author} {\bibfnamefont {A.~I.}\ \bibnamefont
  {Krylov}},\ }\bibfield  {title} {\enquote {\bibinfo {title}
  {Equation-of-motion coupled-cluster methods for open-shell and electronically
  excited species: The hitchhiker's guide to fock space},}\ }\href@noop {}
  {\bibfield  {journal} {\bibinfo  {journal} {Annu. Rev. Phys. Chem.}\ }\textbf
  {\bibinfo {volume} {59}},\ \bibinfo {pages} {433--462} (\bibinfo {year}
  {2008})}\BibitemShut {NoStop}%
\bibitem [{\citenamefont {Stefanucci}\ and\ \citenamefont
  {Van~Leeuwen}(2013)}]{stefanucci2013non}%
  \BibitemOpen
  \bibfield  {author} {\bibinfo {author} {\bibfnamefont {G.}~\bibnamefont
  {Stefanucci}}\ and\ \bibinfo {author} {\bibfnamefont {R.}~\bibnamefont
  {Van~Leeuwen}},\ }\href@noop {} {\emph {\bibinfo {title} {Nonequilibrium
  Many-Body Theory of Quantum Systems: a Modern Introduction}}}\ (\bibinfo
  {publisher} {Cambridge University Press},\ \bibinfo {year}
  {2013})\BibitemShut {NoStop}%
\bibitem [{\citenamefont {Economou}(2006)}]{economou2006green}%
  \BibitemOpen
  \bibfield  {author} {\bibinfo {author} {\bibfnamefont {E.~N.}\ \bibnamefont
  {Economou}},\ }\href@noop {} {\emph {\bibinfo {title} {Green's functions in
  quantum physics}}},\ Vol.~\bibinfo {volume} {7}\ (\bibinfo  {publisher}
  {Springer Science \& Business Media},\ \bibinfo {year} {2006})\BibitemShut
  {NoStop}%
\bibitem [{\citenamefont {Kadanoff}\ and\ \citenamefont
  {Baym}(2018)}]{kadanoff2018quantum}%
  \BibitemOpen
  \bibfield  {author} {\bibinfo {author} {\bibfnamefont {L.~P.}\ \bibnamefont
  {Kadanoff}}\ and\ \bibinfo {author} {\bibfnamefont {G.}~\bibnamefont
  {Baym}},\ }\href@noop {} {\emph {\bibinfo {title} {Quantum Statistical
  Mechanics: Green’s Function Methods in Equilibrium and Nonequilibrium
  Problems}}}\ (\bibinfo  {publisher} {CRC Press},\ \bibinfo {year}
  {2018})\BibitemShut {NoStop}%
\bibitem [{\citenamefont {Hybertsen}\ and\ \citenamefont
  {Louie}(1986)}]{hybertsen1986electron}%
  \BibitemOpen
  \bibfield  {author} {\bibinfo {author} {\bibfnamefont {M.~S.}\ \bibnamefont
  {Hybertsen}}\ and\ \bibinfo {author} {\bibfnamefont {S.~G.}\ \bibnamefont
  {Louie}},\ }\bibfield  {title} {\enquote {\bibinfo {title} {Electron
  correlation in semiconductors and insulators: Band gaps and quasiparticle
  energies},}\ }\href@noop {} {\bibfield  {journal} {\bibinfo  {journal} {Phys.
  Rev. B}\ }\textbf {\bibinfo {volume} {34}},\ \bibinfo {pages} {5390}
  (\bibinfo {year} {1986})}\BibitemShut {NoStop}%
\bibitem [{\citenamefont {Aryasetiawan}\ and\ \citenamefont
  {Gunnarsson}(1998)}]{aryasetiawan1998gw}%
  \BibitemOpen
  \bibfield  {author} {\bibinfo {author} {\bibfnamefont {F.}~\bibnamefont
  {Aryasetiawan}}\ and\ \bibinfo {author} {\bibfnamefont {O.}~\bibnamefont
  {Gunnarsson}},\ }\bibfield  {title} {\enquote {\bibinfo {title} {The gw
  method},}\ }\href@noop {} {\bibfield  {journal} {\bibinfo  {journal} {Rep.
  Prog. Phys.}\ }\textbf {\bibinfo {volume} {61}},\ \bibinfo {pages} {237}
  (\bibinfo {year} {1998})}\BibitemShut {NoStop}%
\bibitem [{\citenamefont {van Schilfgaarde}, \citenamefont {Kotani},\ and\
  \citenamefont {Faleev}(2006)}]{van2006quasiparticle}%
  \BibitemOpen
  \bibfield  {author} {\bibinfo {author} {\bibfnamefont {M.}~\bibnamefont {van
  Schilfgaarde}}, \bibinfo {author} {\bibfnamefont {T.}~\bibnamefont
  {Kotani}},\ and\ \bibinfo {author} {\bibfnamefont {S.}~\bibnamefont
  {Faleev}},\ }\bibfield  {title} {\enquote {\bibinfo {title} {Quasiparticle
  self-consistent gw theory},}\ }\href@noop {} {\bibfield  {journal} {\bibinfo
  {journal} {Phys. Rev. Lett.}\ }\textbf {\bibinfo {volume} {96}},\ \bibinfo
  {pages} {226402} (\bibinfo {year} {2006})}\BibitemShut {NoStop}%
\bibitem [{\citenamefont {Kotani}, \citenamefont {Van~Schilfgaarde},\ and\
  \citenamefont {Faleev}(2007)}]{kotani2007quasiparticle}%
  \BibitemOpen
  \bibfield  {author} {\bibinfo {author} {\bibfnamefont {T.}~\bibnamefont
  {Kotani}}, \bibinfo {author} {\bibfnamefont {M.}~\bibnamefont
  {Van~Schilfgaarde}},\ and\ \bibinfo {author} {\bibfnamefont {S.~V.}\
  \bibnamefont {Faleev}},\ }\bibfield  {title} {\enquote {\bibinfo {title}
  {Quasiparticle self-consistent gw method: A basis for the
  independent-particle approximation},}\ }\href@noop {} {\bibfield  {journal}
  {\bibinfo  {journal} {Phys. Rev. B}\ }\textbf {\bibinfo {volume} {76}},\
  \bibinfo {pages} {165106} (\bibinfo {year} {2007})}\BibitemShut {NoStop}%
\bibitem [{\citenamefont {Golze}, \citenamefont {Dvorak},\ and\ \citenamefont
  {Rinke}(2019)}]{golze2019gw}%
  \BibitemOpen
  \bibfield  {author} {\bibinfo {author} {\bibfnamefont {D.}~\bibnamefont
  {Golze}}, \bibinfo {author} {\bibfnamefont {M.}~\bibnamefont {Dvorak}},\ and\
  \bibinfo {author} {\bibfnamefont {P.}~\bibnamefont {Rinke}},\ }\bibfield
  {title} {\enquote {\bibinfo {title} {The gw compendium: A practical guide to
  theoretical photoemission spectroscopy},}\ }\href@noop {} {\bibfield
  {journal} {\bibinfo  {journal} {Front. Chem.}\ }\textbf {\bibinfo {volume}
  {7}},\ \bibinfo {pages} {377} (\bibinfo {year} {2019})}\BibitemShut {NoStop}%
\bibitem [{\citenamefont {Shishkin}\ and\ \citenamefont
  {Kresse}(2007)}]{shishkin2007self}%
  \BibitemOpen
  \bibfield  {author} {\bibinfo {author} {\bibfnamefont {M.}~\bibnamefont
  {Shishkin}}\ and\ \bibinfo {author} {\bibfnamefont {G.}~\bibnamefont
  {Kresse}},\ }\bibfield  {title} {\enquote {\bibinfo {title} {Self-consistent
  gw calculations for semiconductors and insulators},}\ }\href@noop {}
  {\bibfield  {journal} {\bibinfo  {journal} {Physical Review B}\ }\textbf
  {\bibinfo {volume} {75}},\ \bibinfo {pages} {235102} (\bibinfo {year}
  {2007})}\BibitemShut {NoStop}%
\bibitem [{\citenamefont {Phillips}\ and\ \citenamefont
  {Zgid}(2014)}]{phillips2014communication}%
  \BibitemOpen
  \bibfield  {author} {\bibinfo {author} {\bibfnamefont {J.~J.}\ \bibnamefont
  {Phillips}}\ and\ \bibinfo {author} {\bibfnamefont {D.}~\bibnamefont
  {Zgid}},\ }\bibfield  {title} {\enquote {\bibinfo {title} {Communication: The
  description of strong correlation within self-consistent green's function
  second-order perturbation tteory},}\ }\href@noop {} {\bibfield  {journal}
  {\bibinfo  {journal} {J. Chem. Phys.}\ }\textbf {\bibinfo {volume} {140}},\
  \bibinfo {pages} {241101} (\bibinfo {year} {2014})}\BibitemShut {NoStop}%
\bibitem [{\citenamefont {Pavo{\v{s}}evi{\'c}}\ \emph
  {et~al.}(2017)\citenamefont {Pavo{\v{s}}evi{\'c}}, \citenamefont {Peng},
  \citenamefont {Ortiz},\ and\ \citenamefont
  {Valeev}}]{pavovsevic2017communication}%
  \BibitemOpen
  \bibfield  {author} {\bibinfo {author} {\bibfnamefont {F.}~\bibnamefont
  {Pavo{\v{s}}evi{\'c}}}, \bibinfo {author} {\bibfnamefont {C.}~\bibnamefont
  {Peng}}, \bibinfo {author} {\bibfnamefont {J.}~\bibnamefont {Ortiz}},\ and\
  \bibinfo {author} {\bibfnamefont {E.~F.}\ \bibnamefont {Valeev}},\ }\bibfield
   {title} {\enquote {\bibinfo {title} {Communication: Explicitly correlated
  formalism for second-order single-particle green’s function},}\ }\href@noop
  {} {\bibfield  {journal} {\bibinfo  {journal} {J. Chem. Phys.}\ }\textbf
  {\bibinfo {volume} {147}},\ \bibinfo {pages} {121101} (\bibinfo {year}
  {2017})}\BibitemShut {NoStop}%
\bibitem [{\citenamefont {Dou}\ \emph {et~al.}(2019)\citenamefont {Dou},
  \citenamefont {Takeshita}, \citenamefont {Chen}, \citenamefont {Baer},
  \citenamefont {Neuhauser},\ and\ \citenamefont {Rabani}}]{dou2019stochastic}%
  \BibitemOpen
  \bibfield  {author} {\bibinfo {author} {\bibfnamefont {W.}~\bibnamefont
  {Dou}}, \bibinfo {author} {\bibfnamefont {T.~Y.}\ \bibnamefont {Takeshita}},
  \bibinfo {author} {\bibfnamefont {M.}~\bibnamefont {Chen}}, \bibinfo {author}
  {\bibfnamefont {R.}~\bibnamefont {Baer}}, \bibinfo {author} {\bibfnamefont
  {D.}~\bibnamefont {Neuhauser}},\ and\ \bibinfo {author} {\bibfnamefont
  {E.}~\bibnamefont {Rabani}},\ }\bibfield  {title} {\enquote {\bibinfo {title}
  {Stochastic resolution of identity for real-time second-order green’s
  function: Ionization potential and quasi-particle spectrum},}\ }\href@noop {}
  {\bibfield  {journal} {\bibinfo  {journal} {J. Chem. Theory Comput.}\
  }\textbf {\bibinfo {volume} {15}},\ \bibinfo {pages} {6703--6711} (\bibinfo
  {year} {2019})}\BibitemShut {NoStop}%
\bibitem [{\citenamefont {van Setten}\ \emph {et~al.}(2015)\citenamefont {van
  Setten}, \citenamefont {Caruso}, \citenamefont {Sharifzadeh}, \citenamefont
  {Ren}, \citenamefont {Scheffler}, \citenamefont {Liu}, \citenamefont
  {Lischner}, \citenamefont {Lin}, \citenamefont {Deslippe}, \citenamefont
  {Louie} \emph {et~al.}}]{van2015gw}%
  \BibitemOpen
  \bibfield  {author} {\bibinfo {author} {\bibfnamefont {M.~J.}\ \bibnamefont
  {van Setten}}, \bibinfo {author} {\bibfnamefont {F.}~\bibnamefont {Caruso}},
  \bibinfo {author} {\bibfnamefont {S.}~\bibnamefont {Sharifzadeh}}, \bibinfo
  {author} {\bibfnamefont {X.}~\bibnamefont {Ren}}, \bibinfo {author}
  {\bibfnamefont {M.}~\bibnamefont {Scheffler}}, \bibinfo {author}
  {\bibfnamefont {F.}~\bibnamefont {Liu}}, \bibinfo {author} {\bibfnamefont
  {J.}~\bibnamefont {Lischner}}, \bibinfo {author} {\bibfnamefont
  {L.}~\bibnamefont {Lin}}, \bibinfo {author} {\bibfnamefont {J.~R.}\
  \bibnamefont {Deslippe}}, \bibinfo {author} {\bibfnamefont {S.~G.}\
  \bibnamefont {Louie}}, \emph {et~al.},\ }\bibfield  {title} {\enquote
  {\bibinfo {title} {Gw 100: Benchmarking g 0 w 0 for molecular systems},}\
  }\href@noop {} {\bibfield  {journal} {\bibinfo  {journal} {J. Chem. Theory
  Comput.}\ }\textbf {\bibinfo {volume} {11}},\ \bibinfo {pages} {5665--5687}
  (\bibinfo {year} {2015})}\BibitemShut {NoStop}%
\bibitem [{\citenamefont {Caruso}\ \emph {et~al.}(2016)\citenamefont {Caruso},
  \citenamefont {Dauth}, \citenamefont {Van~Setten},\ and\ \citenamefont
  {Rinke}}]{caruso2016benchmark}%
  \BibitemOpen
  \bibfield  {author} {\bibinfo {author} {\bibfnamefont {F.}~\bibnamefont
  {Caruso}}, \bibinfo {author} {\bibfnamefont {M.}~\bibnamefont {Dauth}},
  \bibinfo {author} {\bibfnamefont {M.~J.}\ \bibnamefont {Van~Setten}},\ and\
  \bibinfo {author} {\bibfnamefont {P.}~\bibnamefont {Rinke}},\ }\bibfield
  {title} {\enquote {\bibinfo {title} {Benchmark of gw approaches for the gw
  100 test set},}\ }\href@noop {} {\bibfield  {journal} {\bibinfo  {journal}
  {J. Chem. Theory Comput.}\ }\textbf {\bibinfo {volume} {12}},\ \bibinfo
  {pages} {5076--5087} (\bibinfo {year} {2016})}\BibitemShut {NoStop}%
\bibitem [{\citenamefont {Rohlfing}\ and\ \citenamefont
  {Louie}(2000)}]{rohlfing2000electron}%
  \BibitemOpen
  \bibfield  {author} {\bibinfo {author} {\bibfnamefont {M.}~\bibnamefont
  {Rohlfing}}\ and\ \bibinfo {author} {\bibfnamefont {S.~G.}\ \bibnamefont
  {Louie}},\ }\bibfield  {title} {\enquote {\bibinfo {title} {Electron-hole
  excitations and optical spectra from first principles},}\ }\href@noop {}
  {\bibfield  {journal} {\bibinfo  {journal} {Phys. Rev. B}\ }\textbf {\bibinfo
  {volume} {62}},\ \bibinfo {pages} {4927} (\bibinfo {year}
  {2000})}\BibitemShut {NoStop}%
\bibitem [{\citenamefont {Attaccalite}, \citenamefont {Gr{\"u}ning},\ and\
  \citenamefont {Marini}(2011)}]{attaccalite2011real}%
  \BibitemOpen
  \bibfield  {author} {\bibinfo {author} {\bibfnamefont {C.}~\bibnamefont
  {Attaccalite}}, \bibinfo {author} {\bibfnamefont {M.}~\bibnamefont
  {Gr{\"u}ning}},\ and\ \bibinfo {author} {\bibfnamefont {A.}~\bibnamefont
  {Marini}},\ }\bibfield  {title} {\enquote {\bibinfo {title} {Real-time
  approach to the optical properties of solids and nanostructures:
  Time-dependent bethe-salpeter equation},}\ }\href@noop {} {\bibfield
  {journal} {\bibinfo  {journal} {Phys. Rev. B}\ }\textbf {\bibinfo {volume}
  {84}},\ \bibinfo {pages} {245110} (\bibinfo {year} {2011})}\BibitemShut
  {NoStop}%
\bibitem [{\citenamefont {Dou}\ \emph {et~al.}(2020)\citenamefont {Dou},
  \citenamefont {Chen}, \citenamefont {Takeshita}, \citenamefont {Baer},
  \citenamefont {Neuhauser},\ and\ \citenamefont {Rabani}}]{dou2020range}%
  \BibitemOpen
  \bibfield  {author} {\bibinfo {author} {\bibfnamefont {W.}~\bibnamefont
  {Dou}}, \bibinfo {author} {\bibfnamefont {M.}~\bibnamefont {Chen}}, \bibinfo
  {author} {\bibfnamefont {T.~Y.}\ \bibnamefont {Takeshita}}, \bibinfo {author}
  {\bibfnamefont {R.}~\bibnamefont {Baer}}, \bibinfo {author} {\bibfnamefont
  {D.}~\bibnamefont {Neuhauser}},\ and\ \bibinfo {author} {\bibfnamefont
  {E.}~\bibnamefont {Rabani}},\ }\bibfield  {title} {\enquote {\bibinfo {title}
  {Range-separated stochastic resolution of identity: Formulation and
  application to second-order green’s function theory},}\ }\href@noop {}
  {\bibfield  {journal} {\bibinfo  {journal} {J. Chem. Phys.}\ }\textbf
  {\bibinfo {volume} {153}},\ \bibinfo {pages} {074113} (\bibinfo {year}
  {2020})}\BibitemShut {NoStop}%
\bibitem [{\citenamefont {Takeshita}\ \emph {et~al.}(2017)\citenamefont
  {Takeshita}, \citenamefont {de~Jong}, \citenamefont {Neuhauser},
  \citenamefont {Baer},\ and\ \citenamefont
  {Rabani}}]{takeshita2017stochastic}%
  \BibitemOpen
  \bibfield  {author} {\bibinfo {author} {\bibfnamefont {T.~Y.}\ \bibnamefont
  {Takeshita}}, \bibinfo {author} {\bibfnamefont {W.~A.}\ \bibnamefont
  {de~Jong}}, \bibinfo {author} {\bibfnamefont {D.}~\bibnamefont {Neuhauser}},
  \bibinfo {author} {\bibfnamefont {R.}~\bibnamefont {Baer}},\ and\ \bibinfo
  {author} {\bibfnamefont {E.}~\bibnamefont {Rabani}},\ }\bibfield  {title}
  {\enquote {\bibinfo {title} {Stochastic formulation of the resolution of
  identity: Application to second order m{\o}ller--plesset perturbation
  theory},}\ }\href@noop {} {\bibfield  {journal} {\bibinfo  {journal} {J.
  Chem. Theory Comput.}\ }\textbf {\bibinfo {volume} {13}},\ \bibinfo {pages}
  {4605--4610} (\bibinfo {year} {2017})}\BibitemShut {NoStop}%
\bibitem [{\citenamefont {Kutz}\ \emph {et~al.}(2016)\citenamefont {Kutz},
  \citenamefont {Brunton}, \citenamefont {Brunton},\ and\ \citenamefont
  {Proctor}}]{kutz2016dynamic}%
  \BibitemOpen
  \bibfield  {author} {\bibinfo {author} {\bibfnamefont {J.~N.}\ \bibnamefont
  {Kutz}}, \bibinfo {author} {\bibfnamefont {S.~L.}\ \bibnamefont {Brunton}},
  \bibinfo {author} {\bibfnamefont {B.~W.}\ \bibnamefont {Brunton}},\ and\
  \bibinfo {author} {\bibfnamefont {J.~L.}\ \bibnamefont {Proctor}},\
  }\href@noop {} {\emph {\bibinfo {title} {Dynamic mode decomposition:
  data-driven modeling of complex systems}}}\ (\bibinfo  {publisher} {SIAM},\
  \bibinfo {year} {2016})\BibitemShut {NoStop}%
\bibitem [{\citenamefont {Yin}\ \emph {et~al.}(2023)\citenamefont {Yin},
  \citenamefont {Chan}, \citenamefont {Felipe}, \citenamefont {Qiu},
  \citenamefont {Yang},\ and\ \citenamefont {Louie}}]{yin2023analyzing}%
  \BibitemOpen
  \bibfield  {author} {\bibinfo {author} {\bibfnamefont {J.}~\bibnamefont
  {Yin}}, \bibinfo {author} {\bibfnamefont {Y.-h.}\ \bibnamefont {Chan}},
  \bibinfo {author} {\bibfnamefont {H.}~\bibnamefont {Felipe}}, \bibinfo
  {author} {\bibfnamefont {D.~Y.}\ \bibnamefont {Qiu}}, \bibinfo {author}
  {\bibfnamefont {C.}~\bibnamefont {Yang}},\ and\ \bibinfo {author}
  {\bibfnamefont {S.~G.}\ \bibnamefont {Louie}},\ }\bibfield  {title} {\enquote
  {\bibinfo {title} {Analyzing and predicting non-equilibrium many-body
  dynamics via dynamic mode decomposition},}\ }\href@noop {} {\bibfield
  {journal} {\bibinfo  {journal} {J. Comput. Phys.}\ ,\ \bibinfo {pages}
  {111909}} (\bibinfo {year} {2023})}\BibitemShut {NoStop}%
\bibitem [{\citenamefont {Yin}\ \emph {et~al.}(2022)\citenamefont {Yin},
  \citenamefont {Chan}, \citenamefont {Felipe}, \citenamefont {Qiu},
  \citenamefont {Louie},\ and\ \citenamefont {Yang}}]{yin2022using}%
  \BibitemOpen
  \bibfield  {author} {\bibinfo {author} {\bibfnamefont {J.}~\bibnamefont
  {Yin}}, \bibinfo {author} {\bibfnamefont {Y.-h.}\ \bibnamefont {Chan}},
  \bibinfo {author} {\bibfnamefont {H.}~\bibnamefont {Felipe}}, \bibinfo
  {author} {\bibfnamefont {D.~Y.}\ \bibnamefont {Qiu}}, \bibinfo {author}
  {\bibfnamefont {S.~G.}\ \bibnamefont {Louie}},\ and\ \bibinfo {author}
  {\bibfnamefont {C.}~\bibnamefont {Yang}},\ }\bibfield  {title} {\enquote
  {\bibinfo {title} {Using dynamic mode decomposition to predict the dynamics
  of a two-time non-equilibrium green’s function},}\ }\href@noop {}
  {\bibfield  {journal} {\bibinfo  {journal} {J. Comput. Sci.}\ }\textbf
  {\bibinfo {volume} {64}},\ \bibinfo {pages} {101843} (\bibinfo {year}
  {2022})}\BibitemShut {NoStop}%
\bibitem [{\citenamefont {Reeves}\ \emph {et~al.}(2023)\citenamefont {Reeves},
  \citenamefont {Yin}, \citenamefont {Zhu}, \citenamefont {Ibrahim},
  \citenamefont {Yang},\ and\ \citenamefont {Vl\ifmmode~\check{c}\else
  \v{c}\fi{}ek}}]{reeves2022dynamic}%
  \BibitemOpen
  \bibfield  {author} {\bibinfo {author} {\bibfnamefont {C.~C.}\ \bibnamefont
  {Reeves}}, \bibinfo {author} {\bibfnamefont {J.}~\bibnamefont {Yin}},
  \bibinfo {author} {\bibfnamefont {Y.}~\bibnamefont {Zhu}}, \bibinfo {author}
  {\bibfnamefont {K.~Z.}\ \bibnamefont {Ibrahim}}, \bibinfo {author}
  {\bibfnamefont {C.}~\bibnamefont {Yang}},\ and\ \bibinfo {author}
  {\bibfnamefont {V.}~\bibnamefont {Vl\ifmmode~\check{c}\else \v{c}\fi{}ek}},\
  }\bibfield  {title} {\enquote {\bibinfo {title} {Dynamic mode decomposition
  for extrapolating nonequilibrium green's-function dynamics},}\ }\href
  {https://doi.org/10.1103/PhysRevB.107.075107} {\bibfield  {journal} {\bibinfo
   {journal} {Phys. Rev. B}\ }\textbf {\bibinfo {volume} {107}},\ \bibinfo
  {pages} {075107} (\bibinfo {year} {2023})}\BibitemShut {NoStop}%
\bibitem [{\citenamefont {Koopman}(1931)}]{koopman1931ham}%
  \BibitemOpen
  \bibfield  {author} {\bibinfo {author} {\bibfnamefont {B.~O.}\ \bibnamefont
  {Koopman}},\ }\bibfield  {title} {\enquote {\bibinfo {title} {Hamiltonian
  systems and transformation in hilbert space},}\ }\href@noop {} {\bibfield
  {journal} {\bibinfo  {journal} {Proc. Natl. Acad. Sci. U.S.A.}\ }\textbf
  {\bibinfo {volume} {17}},\ \bibinfo {pages} {315--318} (\bibinfo {year}
  {1931})}\BibitemShut {NoStop}%
\bibitem [{\citenamefont {Koopman}\ and\ \citenamefont
  {Neumann}(1932)}]{koopman1932dyn}%
  \BibitemOpen
  \bibfield  {author} {\bibinfo {author} {\bibfnamefont {B.~O.}\ \bibnamefont
  {Koopman}}\ and\ \bibinfo {author} {\bibfnamefont {J.~v.}\ \bibnamefont
  {Neumann}},\ }\bibfield  {title} {\enquote {\bibinfo {title} {Dynamical
  systems of continuous spectra},}\ }\href@noop {} {\bibfield  {journal}
  {\bibinfo  {journal} {Proc. Natl. Acad. Sci. U.S.A.}\ }\textbf {\bibinfo
  {volume} {18}},\ \bibinfo {pages} {255--263} (\bibinfo {year}
  {1932})}\BibitemShut {NoStop}%
\bibitem [{\citenamefont {Takeshita}\ \emph {et~al.}(2019)\citenamefont
  {Takeshita}, \citenamefont {Dou}, \citenamefont {Smith}, \citenamefont
  {de~Jong}, \citenamefont {Baer}, \citenamefont {Neuhauser},\ and\
  \citenamefont {Rabani}}]{takeshita2019stochastic}%
  \BibitemOpen
  \bibfield  {author} {\bibinfo {author} {\bibfnamefont {T.~Y.}\ \bibnamefont
  {Takeshita}}, \bibinfo {author} {\bibfnamefont {W.}~\bibnamefont {Dou}},
  \bibinfo {author} {\bibfnamefont {D.~G.}\ \bibnamefont {Smith}}, \bibinfo
  {author} {\bibfnamefont {W.~A.}\ \bibnamefont {de~Jong}}, \bibinfo {author}
  {\bibfnamefont {R.}~\bibnamefont {Baer}}, \bibinfo {author} {\bibfnamefont
  {D.}~\bibnamefont {Neuhauser}},\ and\ \bibinfo {author} {\bibfnamefont
  {E.}~\bibnamefont {Rabani}},\ }\bibfield  {title} {\enquote {\bibinfo {title}
  {Stochastic resolution of identity second-order matsubara green’s function
  theory},}\ }\href@noop {} {\bibfield  {journal} {\bibinfo  {journal} {J.
  Chem. Phys.}\ }\textbf {\bibinfo {volume} {151}},\ \bibinfo {pages} {044114}
  (\bibinfo {year} {2019})}\BibitemShut {NoStop}%
\bibitem [{\citenamefont {Neuhauser}, \citenamefont {Baer},\ and\ \citenamefont
  {Rabani}(2014)}]{neuhauser2014communication}%
  \BibitemOpen
  \bibfield  {author} {\bibinfo {author} {\bibfnamefont {D.}~\bibnamefont
  {Neuhauser}}, \bibinfo {author} {\bibfnamefont {R.}~\bibnamefont {Baer}},\
  and\ \bibinfo {author} {\bibfnamefont {E.}~\bibnamefont {Rabani}},\
  }\bibfield  {title} {\enquote {\bibinfo {title} {Communication: Embedded
  fragment stochastic density functional theory},}\ }\href@noop {} {\bibfield
  {journal} {\bibinfo  {journal} {J. Chem. Phys.}\ }\textbf {\bibinfo {volume}
  {141}},\ \bibinfo {pages} {041102} (\bibinfo {year} {2014})}\BibitemShut
  {NoStop}%
\bibitem [{\citenamefont {Rabani}, \citenamefont {Baer},\ and\ \citenamefont
  {Neuhauser}(2015)}]{rabani2015time}%
  \BibitemOpen
  \bibfield  {author} {\bibinfo {author} {\bibfnamefont {E.}~\bibnamefont
  {Rabani}}, \bibinfo {author} {\bibfnamefont {R.}~\bibnamefont {Baer}},\ and\
  \bibinfo {author} {\bibfnamefont {D.}~\bibnamefont {Neuhauser}},\ }\bibfield
  {title} {\enquote {\bibinfo {title} {Time-dependent stochastic bethe-salpeter
  approach},}\ }\href@noop {} {\bibfield  {journal} {\bibinfo  {journal}
  {Physical Review B}\ }\textbf {\bibinfo {volume} {91}},\ \bibinfo {pages}
  {235302} (\bibinfo {year} {2015})}\BibitemShut {NoStop}%
\end{thebibliography}%
\end{document}